\documentclass[preprint,superscriptaddress,longbibliography,prd,aps,nofootinbib,showkeys, 10pt]{revtex4-2}

\pdfoutput=1

\usepackage{orcidlink}  

\usepackage{keyval,amsfonts, slashed,bm}
\usepackage{graphicx,textcomp}

\usepackage{amsmath,amssymb}
\usepackage{color}
\usepackage[normalem]{ulem}
\usepackage{xcolor}
\DeclareMathOperator{\arccot}{arccot}

\def\be {\begin{equation}}
	\def\ee {\end{equation}}
\def\bea {\begin{eqnarray}}
	\def\eea {\end{eqnarray}}
\def\bc {\begin{center}}
	\def\ec {\end{center}}
\def\nn {\nonumber}

\def\({\left(}
\def\){\right)}
\def\[{\left[}
\def\]{\right]}

\newcommand \Tr{\operatorname{\text{Tr}}}

\definecolor{sinopia}{rgb}{0.8,0.25,0.04}

\definecolor{greenopia}{rgb}{0.3,0.65,0.14}

\begin{document}

\title{Complex heavy-quarkonium potential in an anisotropic collisional quark-gluon plasma}                      

\author{Manas Debnath, \orcidlink{0009-0003-8488-7146}}
	\affiliation{School of Physical Sciences, National Institute of Science Education and Research, An OCC of Homi Bhabha National Institute,\\  Jatni, Khurda 752050, India}
    
	\author{Lata Thakur,\orcidlink{0000-0003-2343-4963}}
	\affiliation{Department of Physics and Institute of Physics and Applied Physics, Yonsei University,\\ Seoul 03722, Korea}

    \author{Najmul Haque,\orcidlink{0000-0001-6448-089X}}
	\email{nhaque@niser.ac.in}
	\affiliation{School of Physical Sciences, National Institute of Science Education and Research, An OCC of Homi Bhabha National Institute,\\  Jatni, Khurda 752050, India}
	%

\begin{abstract}
We compute the complex heavy-quark potential in an anisotropic quark-gluon plasma (QGP) using kinetic theory with a Bhatnagar-Gross-Krook collision kernel. By incorporating momentum anisotropy and a finite collision rate into the medium dielectric response, we derive both the real and the imaginary parts of the in-medium potential. The real part of the inverse dielectric function is obtained from the retarded/advanced gluon propagator, while the imaginary part is determined from the Feynman (symmetric) propagator. We find that collisions have only a minimal impact on the real part of the potential, suggesting a similarly weak effect on the binding energy. In an anisotropic plasma, the Weibel instability induces a pinch singularity that can render the imaginary part of the potential ill-defined; we show that sufficiently large collision rates regularize this singularity, yielding a well-defined imaginary potential in the corresponding region of parameter space. In this well-defined regime, collisions significantly enhance the magnitude of the imaginary part and modify the effect of anisotropy. This enhancement leads to larger quarkonium thermal widths and dissociation rates in a nonequilibrium QGP, providing further insight into quarkonium suppression mechanisms.
\end{abstract}

	
	
	\maketitle
	
	\bigskip
	
\section{Introduction}\label{Int}	
The study of the quark-gluon plasma (QGP) remains a primary objective of high-energy nuclear physics, driven by the experimental programs at the Relativistic Heavy Ion Collider (RHIC) at Brookhaven National Laboratory (BNL) and the ongoing Large Hadron Collider (LHC) at CERN. These experimental programs aim to characterize the properties of the QGP, which is a deconfined state of matter expected to emerge at extremely high temperatures. In this regime, the QGP exhibits the behavior of a weakly interacting system, a feature that can be theoretically investigated using hard thermal loop (HTL) resummation techniques~\cite{Weldon:1982aq,Braaten:1989mz,Frenkel:1989br,Braaten:1991gm,Haque:2014rua,Haque:2024gva}.

Heavy quarkonium suppression is widely regarded as a signature of QGP formation~\cite{Matsui:1986dk}. In a vacuum, quarkonium states are well-described by nonrelativistic potential models characterized by Coulombic behavior at short distances and linear confinement at larger separations~\cite{Eichten:1974af,Eichten:1978tg,Eichten:1979ms,Buchmuller:1980su,Koma:2006si,Brambilla:2004jw}. However, within a hot medium, this potential is significantly modified by Debye screening, leading to the dissociation of bound states above a critical crossover temperature~\cite{Matsui:1986dk}. Since the binding energies of different quarkonium states range from several mega-electron volts (MeV) to a few giga-electron volts (GeV), they are expected to undergo sequential suppression~\cite{Karsch:2005nk}. In this process, weakly bound states melt near the transition temperature while more tightly bound states persist until higher temperatures. This phenomenon has been observed at RHIC and the LHC, where higher-mass states of bottomonia and charmonia~\cite{CMS:2011all,CMS:2012gvv} exhibit systematically stronger suppression.

While earlier studies of heavy quarkonium typically assumed an isotropic medium~\cite{Agotiya:2008ie}, it is now understood that the early stages of heavy-ion collisions are characterized by momentum-space anisotropies resulting from rapid longitudinal expansion~\cite{Romatschke:2003ms,Romatschke:2004jh,Mrowczynski:2004kv}. Noncentral collisions, in particular, generate anisotropic momentum distributions that lead to differential pressures and kinetic instabilities~\cite{Mrowczynski:2016etf}. Such instabilities significantly influence the system's evolution and accelerate isotropization~\cite{Romatschke:2006wg,Mrowczynski:2016etf}. Furthermore, these anisotropies modify the collective gluon modes~\cite{Romatschke:2003ms,Dumitru:2009fy,Karmakar:2022one}, which in turn alter the heavy-quark potential~\cite{Dumitru:2007hy,Burnier:2009yu,Dumitru:2009ni,Dumitru:2009fy,Thakur:2012eb,Dong:2022mbo,Ghosh:2022sxi,Carrington:2024rxw}. In addition to anisotropy, collisions among plasma constituents become essential to consider as the system cools~\cite{Zhao:2023mrz}, even though they are often ignored at very high temperatures due to the separation of timescales between hard and soft momentum exchanges.

The study of gluon collective modes in collisional QGP has evolved significantly. Initial work utilized the Bhatnagar-Gross-Krook (BGK) collision kernel for isotropic thermal media~\cite{Carrington:2003je}, which was later extended to anisotropic plasmas for modes propagating parallel to the anisotropy vector~\cite{Schenke:2006xu}. Subsequent research considered modes propagating in all directions relative to a small anisotropy vector within an effective quasiparticle model~\cite{Kumar:2017bja}. Recently, Ref.~\cite{Zhao:2023mrz} provided a more general framework by incorporating the full angular dependence of gluon propagation across arbitrary anisotropy strengths.

Despite these advancements, most investigations into quarkonium properties in anisotropic QGP have remained restricted to collisionless scenarios~\cite{Dumitru:2007hy,Dumitru:2009fy,Thakur:2012eb,Thakur:2013nia,Nopoush:2017zbu,Carrington:1997sq}. In this work, we bridge this gap by systematically deriving the complex heavy-quark potential in a momentum-anisotropic QGP with finite collisional rates. While the collective modes derived in Ref.~\cite{Zhao:2023mrz} allow for the determination of the real part of the potential, evaluating the imaginary component requires the Feynman (symmetric) propagator within the real-time formalism. Since the derivation of this propagator for an anisotropic collisional medium is currently absent in the literature, we establish it here for the first time and utilize it to compute the imaginary potential, accounting for both general anisotropy and the BGK collision kernel.

Calculating the imaginary potential in this context necessitates careful treatment of the integrand. In the collisionless limit with small anisotropy ($\xi$), the imaginary potential can be determined through a first-order expansion in $\xi$~\cite{Thakur:2013nia}. However, for general anisotropy strengths, performing such an expansion is not appropriate, and one must evaluate the full expression. In this case, a singularity appears in the denominator of the integrand, which is associated with the Weibel instability~\cite{Nopoush:2017zbu}. We address this challenge by identifying parameter regimes in which finite collisional effects regularize the singularity, thereby ensuring a well-defined imaginary potential. Our analysis elucidates how the interplay of collisions and anisotropy influences quarkonium thermal widths and dissociation rates, providing critical insights into the suppression mechanisms relevant to relativistic heavy-ion collisions.

Beyond anisotropies and collisions, quarkonium properties are also influenced by other nontrivial environmental factors. These include the strong magnetic fields generated in noncentral collisions~\cite{Sebastian:2023tlw, Singh:2017nfa}, the relative motion between the medium and the quarkonium bound state~\cite{Thakur:2016cki, Sebastian:2022sga}, bulk viscous effects~\cite{Du:2016wdx, Thakur:2020ifi, Thakur:2021vbo}, and modifications arising from the Gribov–Zwanziger action~\cite{Debnath:2023dhs}.

The paper is organized as follows. Section~\ref{sec:prop} discusses the gluon propagator in a collisional anisotropic QGP. Section~\ref{sec:eps} details the dielectric permittivity of the medium. In Sec.~\ref{sec:pot}, we derive the complex heavy-quark potential and analyze the impacts of collisions and anisotropy. Sec.~\ref{sec:dw} focuses on the thermal width of the charmonium ground state. Finally, we summarize our results and discuss their implications in Sec.~\ref{sec:summary}.
\section{Gluon propagator in a collisional anisotropic medium}
\label{sec:prop}
In this section, we discuss the gluon self-energy and propagator in the presence of a collisional anisotropic QGP medium, which we later use to compute the permittivity of the medium and eventually use that to determine the in-medium heavy quark-antiquark potential.
    
The momentum-space anisotropic correction is introduced by deforming the distribution functions of constituent particles~\cite{Romatschke:2003ms}. The   deformed distribution function is parametrized in the following spheroidal form 
	\begin{align}
    \label{f_dist}
		f(\bm{k}) & =f_{\text {iso }}\left(\frac{1}{\lambda} \sqrt{k_T^2+(1+\xi) k_L^2}\right) \nonumber\\
		& =f_{\text {iso }}\left(\frac{1}{\lambda} \sqrt{\bm{k}^2+\xi(\bm{k} \cdot \bm{n})^2}\right),
	\end{align}
where $f(\bm{k})$ is derived from $f_{\rm iso}(k)$ by filtering out particles with a large momentum component along the anisotropic direction, $\bm{n}$~\cite{Romatschke:2003ms}. 
Here, $ k_T $ and $ k_L $ denote the components of momentum transverse and longitudinal to a given anisotropy axis $\bm{n}$ with $k=|\bm{k}|=\sqrt{k_T^2+k_L^2}$. Here, in Eq.~\eqref{f_dist}, the parameter $ \xi $ quantitatively measures the strength of momentum-space anisotropy. 
The scale $\lambda$ denotes the characteristic hard momentum scale of the plasma particles. It coincides with the temperature only in the thermal-equilibrium limit. For the spheroidal deformation considered here, the system is characterized by a single anisotropy direction. As a result, the modes of the gluon polarization tensor depend on the propagation angle relative to $ \bm{n}$. We first calculate the gluon self-energy and propagators in a collisional anisotropic QGP medium. Subsequently, we determine the inverse dielectric permittivity of the anisotropic medium with collision, which will be used to compute the real and imaginary parts of the quarkonium static potential. 
\subsection{Gluon self-energy and resummed propagator}
In the collisionless case, the gluon self-energy is typically obtained by the HTL resummation technique from the Feynman diagrams.
However, for the collisional QGP, the gluon self-energy is usually obtained using kinetic theory~\cite{Carrington:2003je,Schenke:2006xu,Zhao:2023mrz}.
Within the framework of kinetic theory, the phase-space distribution of partons (i.e. gluons, quarks, and antiquarks) in the QCD plasma is characterized using gauge-covariant Wigner functions, $W^i(\bm{k},X)$ with $i\in \{g,q,\Bar{q}\}$ and $\bm{k}$ and $X$ denote the three-momentum and four-position, respectively. These particles are treated as massless and satisfy the mass-shell constraint. To obtain the linearized transport equation, we expand $W^i(\bm{k},X)=W^i(\bm{k})+\delta W^i(\bm{k},X)$. After applying a gauge-covariant gradient expansion, the generalized Boltzmann equation governing quarks and gluons takes the form
\begin{align}
\label{boltzmann_eq}
V\cdot \partial_X\delta f^i_a (\bm{k},X) + g \theta_i\, V_{\mu} F^{\mu\nu}_a \partial_{\nu}^{(K)}f^i(\bm{k})=\mathcal{C}^i_a(\bm{k},X) ,
\end{align}
where $V=(1,\bm{v})$ with $\bm{v}\equiv\bm{k}/k$ and $\theta_g=\theta_q=1, \theta_{\Bar{q}}=-1$. 
In Eq.~\eqref{boltzmann_eq}, the weak-coupling limit is taken, and higher-order terms of coupling $g$ are neglected. Thus, the field tensor becomes $F^{\mu\nu}=\partial_{\mu}A_{\nu}-\partial_{\nu}A_{\mu}+\mathcal{O}(g)$. In this limit, the theory becomes effectively Abelian, and one may consider that there is no coupling between different color channels. The distribution function $f^i(\bm{k)}$ and its color fluctuations $\delta f^i_a(\bm{k},X)$ are related via the Wigner function as
\begin{align}
f^{q/\Bar{q}}(\bm{k})= \frac{1}{N_c} \Tr [W^{q/\Bar{q}}(\bm{k},X)], \qquad f^{g}(\bm{k})= \frac{1}{N_c^2-1} \Tr [W^{g}(\bm{k},X)],
\end{align}
and 
\begin{align}
\delta f^{q/\Bar{q}}_a(\bm{k},X)= 2 \Tr[t_a\, \delta W^{q/\Bar{q}}(\bm{k},X)], \qquad \delta f^g_a(\bm{k},X)= \frac{1}{N_c} \Tr [T_a\, \delta W^g(\bm{k},X)],
\end{align}
where $t_a$ are the $SU(N_c)$ generators in the fundamental representation, and $T_a$ are the corresponding adjoint-representation generators.
Now, for the collisionless case, the right-hand side of Eq.~\eqref{boltzmann_eq} is simply zero, that is, $\mathcal{C}^i_a(\bm{k},X)=0$. When collisions are included, $\mathcal{C}^i_a(\bm{k},X)$ is given by a BGK-type collision kernel
\bea\label{collision_term}
\mathcal{C}^i_a(\bm{k},X) &=&-\nu \, \bigg[f^i_a(\bm{k},X) - \frac{N^i_a(X)}{N^i_{eq}} f^i_{eq}(k) \bigg],
\eea
with $f^i_a(\bm{k},X)=f^i(\bm{k})+ \delta f^i_a(\bm{k},X)$, and 
\bea
N^i_a(X)= \int_{\bm{k}} f^i_a(\bm{k},X),  \qquad N^i_{eq}= \int_{\bm{k}} f^i_{eq}(k).
\eea
The BGK-type collision term in Eq.~\eqref{collision_term} characterizes the equilibration of the system through collisions, occurring over a timescale inversely proportional to $\nu$.
In this work, the collision rate $\nu$ is treated as a constant free parameter, independent of particle momentum and species. An advantage of the BGK kernel is that it ensures local particle-number conservation
; that is,
 \be
\int_{\bm{k}}\mathcal{C}^i_a(\bm{k},X)=0.
 \ee
 In Ref.~\cite{Schenke:2006xu}, it is estimated that $\nu$ lies in the range $\nu \sim 0.1 m_D-0.2 m_D$, but it is possible that the collision rate could even be larger than this~\footnote{For the imaginary part of the potential, we restrict to $\nu\in[0.25\,m_D,0.45\,m_D]$, where the imaginary part is well defined, because a singular region persists up to $\nu\sim 0.24m_D$ near $\xi\sim 10$. See Fig.~\ref{fig:singularity}}. The induced current for each color channel is given by,
 \bea
J^{\mu}_{ind,a}=g\int_{\bm{k}}V^{\mu}\{2N_c\, \delta f^g_a(k,X)+N_f[\delta f^{q}_a(k,X)-\delta f^{\Bar{q}}_a(k,X)]\}.
\label{J_ind}
 \eea
The gluon self-energy for the collisional case can be obtained following Refs.~\cite{Zhao:2023mrz, Schenke:2006xu} as
\begin{eqnarray}
\label{self energy}
\Pi^{\mu \nu}_{ab}(P)&=&\frac{\delta J^{\mu}_{ind,a}(P)}{\delta A^b_{\nu}(P)}\nn\\
&=& g^2 \delta_{ab}\int_{\bm{k}} V^{\mu}\partial_l^{\bm{k}} f(\bm{k})\frac{g^{l \nu}(\hat{\omega}-\bm{\hat{p}}\cdot \bm{v})-\hat{P}^l V^{\nu}}{\hat{\omega}-\bm{\hat{p}}\cdot \bm{v} + i \hat{\nu}}+ g^2 \delta_{ab} (i \hat{\nu}) \int \frac{d\Omega}{4\pi}\frac{V^{\mu}}{\hat{\omega}-\bm{\hat{p}}\cdot \bm{v} + i \hat{\nu}}  \nonumber \\
&\times& \int_{\bm{k}^{\prime}} \partial_l^{(\bm{k}^{\prime})} f(\bm{k}^{\prime})\,\frac{g^{l \nu}(\hat{\omega}-\bm{\hat{p}}\cdot \bm{v}^{\prime})-\hat{P}^l V^{\prime \nu}}{\hat{\omega}-\bm{\hat{p}}\cdot \bm{v}^{\prime} + i \hat{\nu}} \mathcal{W}^{-1}(\hat{\omega},\hat{\nu}),
\end{eqnarray}
where 
$\int_{\bm{k}}\equiv \int \, d^3\bm{k}/(2\pi)^3$. 
For convenience, we define the dimensionless collision rate $\hat{\nu}=\nu/p$. 
$\hat{P}=(\hat{\omega},\hat{\bm{p}})$ with $\hat{\omega}=\omega/p$ and $\hat{\bm{p}}=\bm{p}/p$. The function $\mathcal{W}(\hat{\omega},\hat{\nu})$ is defined as
 \be
\mathcal{W}(\hat{\omega},\hat{\nu})=1- \frac{i \hat{\nu}}{2}\int_{-1}^{1} \frac{dy}{\hat{\omega}-y+i \hat{\nu}}=1-\frac{i\hat{\nu}}{2}\ln \frac{\hat{\omega}+i\hat{\nu}+1}{\hat{\omega}+i\hat{\nu}-1}.
\ee
In  Eq.~\eqref{self energy}, $f(\bm{k})=N_f[f^q(\bm{k})+f^{\Bar{q}}(\bm{k})] +N_c f^g(\bm{k})$, where the distribution function of partons is anisotropic as defined in Eq.~\eqref{f_dist}. The gluon self-energy still satisfies the transversality condition even after the introduction of the collisional term. The spatial component of gluon self-energy can be obtained after substituting Eq.~\eqref{f_dist} into Eq.~\eqref{self energy} as
\begin{eqnarray}
\label{selfenergy2}
\Pi^{ij}(P)&=&m_D^2 \int \frac{d\Omega}{4\pi} v^i \frac{v^l + \xi (\bm{v}\cdot\bm{n})n^l}{(1+ \xi (\bm{v}\cdot\bm{n})^2)^2}\frac{\delta^{lj}(\hat{\omega}-\bm{\hat{p}}\cdot\bm{v})+\hat{p}^l v^j}{\hat{\omega}-\hat{\bm{p}}\cdot \bm{v}+i \hat{\nu}} + i\hat{\nu} m_D^2 \int\frac{d\Omega'}{4\pi} \frac{v'^i}{\hat{\omega}-\hat{\bm{p}}\cdot\bm{v'}+i \hat{\nu}}\nn \\
&\times& \int \frac{d\Omega}{4\pi} \frac{v^l + \xi (\bm{v}\cdot\bm{n})n^l}{(1+ \xi (\bm{v}\cdot\bm{n})^2)^2} \frac{\delta^{lj}(\hat{\omega}-\bm{\hat{p}}\cdot\bm{v})+\hat{p}^l v^j}{\hat{\omega}-\hat{\bm{p}}\cdot \bm{v}+i \hat{\nu}} \, \mathcal{W}(\hat{\omega},\hat{\nu})^{-1},
\label{Pi^ij}
\end{eqnarray}
where 
\be\label{debyemass}
m_D^2=- \frac{g^2}{2\pi^2}\int_0^{\infty}dk \,\, k^2 \frac{d f_{iso}(k)}{dk}.
\ee
The gluon self-energy given in Eq.~\eqref{Pi^ij} is asymmetric in the Lorentz indices due to the presence of a collision term of BGK-type. This asymmetry is also seen in magnetized plasma because of the lack of time-reversal symmetry~\cite{Wang:2021ebh}. 
Therefore, the symmetric tensor basis for anisotropic systems in the collisionless limit is unsuitable for the decomposition in Eq.~\eqref{selfenergy2}. Instead, we adopt a more general tensor basis with five structure functions to decompose the gluon self-energy,
   \be \label{basis}
\Pi^{ij}(P)= \alpha A^{ij}+ \beta B^{ij} + \gamma C^{ij} + \delta D^{ij} + \rho E^{ij},
   \ee
with the tensor basis defined as
\bea
\begin{aligned}
A^{\mu \nu} & =-\eta^{\mu \nu}+\frac{P^\mu P^\nu}{P^2}+\frac{\tilde{m}^\mu \tilde{m}^\nu}{\tilde{m}^2} \\
B^{\mu \nu} & =-\frac{P^2}{(m \cdot P)^2} \frac{\tilde{m}^\mu \tilde{m}^\nu}{\tilde{m}^2} \\
C^{\mu \nu} & =\frac{\tilde{m}^2 P^2}{\tilde{m}^2 P^2+(n \cdot P)^2}\left[\tilde{n}^\mu \tilde{n}^\nu-\frac{\tilde{m} \cdot \tilde{n}}{\tilde{m}^2}\left(\tilde{m}^\mu \tilde{n}^\nu+\tilde{m}^\nu \tilde{n}^\mu\right)+\frac{(\tilde{m} \cdot \tilde{n})^2}{\tilde{m}^4} \tilde{m}^\mu \tilde{m}^\nu\right] \\
D^{\mu \nu} & = \frac{\tilde{m} \cdot \tilde{n}}{\tilde{m}^2} \tilde{m}^\mu \tilde{m}^\nu-\tilde{n}^\mu \tilde{m}^\nu\\
E^{\mu \nu} & =\frac{\tilde{m} \cdot \tilde{n}}{\tilde{m}^2} \tilde{m}^\mu \tilde{m}^\nu-\tilde{m}^\mu \tilde{n}^\nu.
\end{aligned}
\label{tensor}
\eea
Here, $m^\mu$ is the heat-bath four-velocity, which in the local rest frame is given by $m^\mu= (1,0,0,0)$, and
$$
\tilde{m}^\mu=m^\mu-\frac{m \cdot P}{P^2} P^\mu
$$
is the component of $m^\mu$ orthogonal to $p^\mu$. The direction of anisotropy in momentum space is determined by the vector
$$
n^\mu=(0, \bm{n})
$$
where $\bm{n}$ is a three-dimensional unit vector. Likewise, 
$$\tilde{n}^\mu=n^\mu-\frac{n \cdot P}{P^2} P^\mu$$
is the component of $n^\mu$ orthogonal to $P^\mu$. The gluon dressed propagators can also be expanded in terms of the tensors defined in Eq.~\eqref{tensor}.

Using the orthonormal properties of the tensor basis, the five structure functions can be determined from Eq.~\eqref{basis} based on the following contraction as
\begin{subequations}
\label{struc_func}
\begin{align}
\hat{p}^i\Pi^{ij}\hat{p}^j&=\beta,\\
\quad \Tilde{n}^i \Pi^{ij}\Tilde{n}^j &=\Tilde{n}^2 (\alpha+\gamma ),\\
\Tilde{n}^i \Pi^{ij}\hat{p}^j&= \Tilde{n}^2 \delta, \\  \hat{p}^i \Pi^{ij}\Tilde{n}^j &=\Tilde{n}^2 \rho,\\ \text{Tr}\Pi^{ij}&= 2 \alpha + \beta + \gamma . 
\end{align}
\end{subequations}
Using gluon self-energy expression, the inverse of the resummed gluon propagator $\Delta^{ij}(P)$ can be computed in temporal axial gauge (TAG)using the Dyson-Schwinger equation as
   \bea
 (\Delta^{-1})^{ij}(P)&=& (p^2-\omega^2 + \alpha)A^{ij}+ (\beta- \omega^2)B^{ij} 
 +\gamma C^{ij}+ \delta D^{ij}+ \rho E^{ij}.
    \eea    
Upon inversion, the resummed gluon propagator becomes
   \bea\label{gl_prop}
\Delta^{ij}(P)&=\Delta_A [A^{ij}-C^{ij}]+ \Delta_G [(p^2-\omega^2+\alpha + \gamma)B^{ij} 
+(\beta-\omega^2)C^{ij}-\delta D^{ij}- \rho E^{ij}],
   \eea
   with 
\bea
\Delta_A&=& \frac{1}{p^2-\omega^2+\alpha}, \nn\\
   \Delta_G &=& \frac{1}{(p^2- \omega^2 + \alpha + \gamma )(\beta- \omega^2)-  \delta \rho p_\perp^2/p^2}.
\eea
Here, $p_\perp = p\sqrt{1-t^2} = |\bm{p}-(\bm{p\cdot n})\bm{n}|$ is the component of the gluon momentum transverse to the anisotropy direction.
The gluon propagator in Eq.~\eqref{gl_prop} can be rewritten as
 \bea\label{gl_prop_gen}
\Delta^{ij}(P)&=\alpha'A^{ij}+\beta' B^{ij}+\gamma' C^{ij}+\delta' D^{ij}+\rho' E^{ij},
   \eea
   with 
\begin{subequations}
\begin{align}
   \alpha'&=\Delta_A,\\
   \beta'&=\Delta_G (p^2-\omega^2+\alpha+\gamma),\\
   \gamma'&=\Delta_G (\beta-\omega^2)-\Delta_A,\\
   \delta'&=-\Delta_G\,\delta,\\
   \rho'&=-\Delta_G\,\rho. 
\end{align}
\end{subequations}
The dispersion relation can be determined by finding the poles of the propagator $\Delta^{ij}(P)$, namely,
   \bea
   \Delta^{-1}_A&=& p^2-\omega^2+\alpha=0, \nonumber\\
   \Delta^{-1}_G &=& (p^2- \omega^2 + \alpha + \gamma )(\beta- \omega^2)- \Tilde{n}^2 \delta \rho=(\omega^2 - \Omega_{+}^2)(\omega^2 - \Omega_{-}^2)=0,
   \eea
   where
\be
2\Omega^2_{\pm}=\bar{\Omega}^2 \pm \sqrt{\bar{\Omega}^4- 4[(\alpha+\gamma+p^2)\beta-\Tilde{n}^2\delta \rho]},
\ee
   with
\be
 \bar{\Omega}^2=\alpha+\beta+\gamma+p^2.
\ee
From Eq.~\eqref{struc_func}, the structure functions~\cite{Zhao:2023mrz} are obtained in Appendix~\ref{appen_A} as
 \begin{subequations}
\begin{eqnarray}
        \alpha(\xi, \hat{\omega},t,\hat{\nu})&=& \frac{m_D^2}{2(1-t^2)}\left[\hat{\omega}zf_0(\xi)-\xi t^2 f_1(\xi)+\hat{\omega}(1-t^2-z^2)\mathcal{I}^{[0]}(\xi,\hat{\omega},t,\hat{\nu}) \right.\nn \\
        &+&\left. t\left\{\xi (1-t^2)+z(2\hat{\omega}-\xi z)\right\}\mathcal{I}^{[1]}(\xi,\hat{\omega},t,\hat{\nu})-(\hat{\omega}-2 \xi z t^2)\mathcal{I}^{[2]}(\xi,\hat{\omega},t,\hat{\nu})-\xi t \mathcal{I}^{[3]}(\xi,\hat{\omega},t,\hat{\nu})\right],\\
         \beta(\xi,\hat{\omega},t,\hat{\nu})&=&- \frac{\hat{\omega}^2m_D^2}{2\mathcal{W}(\hat{\omega},\hat{\nu})}\Big[f_0 (\xi)-z \mathcal{I}^{[0]}(\xi,\hat{\omega},t,\hat{\nu})-\xi t \mathcal{I}^{[1]}(\xi,\hat{\omega},t,\hat{\nu})\Big], \\
        \gamma(\xi,\hat{\omega},t,\hat{\nu}) &=& \frac{m_D^2}{2(1-t^2)}\Big[\xi (1+t^2)f_1(\xi)-\hat{\omega}z (1+t^2) f_0(\xi)-\hat{\omega}\left\{(1-t^2)-z^2(1+t^2)\right\}\mathcal{I}^{[0]}(\xi,\hat{\omega},t,\hat{\nu}), \nn \\
        &- &t\left\{ 4\hat{\omega}z -2 \xi z^2+\xi(1-t^2)(1+\hat{\omega}z)\right\}\mathcal{I}^{[1]}(\xi,\hat{\omega},t,\hat{\nu}) \nn \\
        &-& \left\{ 4\xi z- 2\hat{\omega}-\xi (1-t^2)(\hat{\omega}+3z)\right\}\mathcal{I}^{[2]}(\xi,\hat{\omega},t,\hat{\nu})+2\xi t \mathcal{I}^{[3]}(\xi,\hat{\omega},t,\hat{\nu})\Big],\\
        \delta(\xi,\hat{\omega},t,\hat{\nu})&=& \frac{\hat{\omega}m_D^2}{2(1-t^2)}\Big[tzf_0(\xi)-tz^2 \mathcal{I}^{[0]}(\xi,\hat{\omega},t,\hat{\nu})+z(1-\xi t^2)\mathcal{I}^{[1]}(\xi,\hat{\omega},t,\hat{\nu})+\xi t \mathcal{I}^{[2]}(\xi,\hat{\omega},t,\hat{\nu})\Big], \\
        \rho(\xi,\hat{\omega},t,\hat{\nu})&=&\frac{\hat{\omega}m_D^2}{2\mathcal{W}(\hat{\omega},\hat{\nu})(1-t^2)}\Big[\hat{\omega} t f_0(\xi)-\hat{\omega}t z \mathcal{I}^{[0]}(\xi,\hat{\omega},t,\hat{\nu})+\left\{\hat{\omega}(1+\xi -\xi t^2)-\xi z\right\} \mathcal{I}^{[1]}(\xi,\hat{\omega},t,\hat{\nu}) \nn \\
        &+&\xi t \mathcal{I}^{[2]}(\xi,\hat{\omega},t,\hat{\nu})\Big],
    \end{eqnarray}
\end{subequations}
 where $z=\hat{\omega}+i\hat{\nu}$. The master integral, $\mathcal{I}^{[n]} (\xi,\hat{\omega},t,\hat{\nu})$ has the form
\bea
\mathcal{I}^{[n]}(\xi,\hat{\omega},t,\hat{\nu}) 
&=&\int_{-1}^{1} dy \, \, \frac{y^n}{(1+ \xi y^2)^2} \int_{0}^{2\pi} \frac{d\phi}{2\pi} \frac{1}{\hat{\omega}-\hat{\bm{p}}\cdot\bm{v}+i\hat{\nu}} \nn \\
&=& \int_{-1}^{1} dy \, \, \frac{y^n}{(1+ \xi y^2)^2} \frac{\text{sign}(\text{Re}[\hat{\omega}]-\hat{p}_z y)}{\sqrt{(\hat{\omega}-\hat{p}_z y + i \hat{\nu})^2- \hat{p}_y^2 (1-y^2)}}, \nn \\
&=& \int_{-1}^{1} dy \, \, \frac{y^n}{(1+ \xi y^2)^2} \frac{\text{sign}(\text{Re}[\hat{\omega}]-t y)}{\sqrt{(\hat{\omega}-t y + i \hat{\nu})^2- (1-t^2) (1-y^2)}},
\eea
\noindent
where $n=0,1,2,3$ and $t=\cos \theta$, with $\theta$ being the angle between $\bm{p}$ and $\bm{n}$. It is assumed that $\bm{p}$ lies in the $y$-$z$ plane, while $\bm{n}$ is parallel to the $z$-axis.  
The frequency $\omega$ is generally complex, and the integration variable $y$ represents the cosine of the polar angle between $\bm{k}$ and $\bm{n}$.
%
The functions $ f_0(\xi)$ and $f_1(\xi)$ are defined as 
 \begin{subequations}
    \begin{eqnarray}
        f_0(\xi)&=& \int_0^1 \frac{2}{(1+\xi y^2)^2}\, dy=\frac{1}{1+\xi}+\frac{\arctan \sqrt{\xi}}{\sqrt{\xi}},\\
        f_1(\xi)&=& \int_0^1 \frac{2 y^2}{(1+\xi y^2)^2}\, dy=-\frac{1}{\xi+\xi^2}+\frac{\arctan \sqrt{\xi}}{\sqrt{\xi^3}}.
    \end{eqnarray}
\end{subequations}
Note that, in the collisionless limit, that is, when the collision parameter $\nu=0$, $\mathcal{W}(\hat{\omega},0)=1$ and $z=\hat{\omega}$. Then,
\bea
\rho (\xi,\hat{\omega},t,0)=\delta (\xi,\hat{\omega},t,0),
\eea
The above relation implies that in the collisionless limit, we have only four structure functions, and the self-energy is symmetric, i.e., $\Pi^{ij}=\Pi^{ji}$.

As the dielectric permittivity of the medium is defined in the static limit ($\omega \rightarrow 0$), we can define various mass scales corresponding to the structure functions as
 \begin{subequations}
    \begin{align}
       \alpha_{R/A}&\approx m_{\alpha}^2 \pm i\hat\omega m_{\alpha, 1}^2+\mathcal{O}(\hat\omega^2) ,\\ 
       \beta_{R/A}&\approx-\hat{\omega}^2\left[ m_{\beta}^2\pm i\hat\omega m_{\beta,1}^2\right]+\mathcal{O}(\hat\omega^4) ,\\
       \gamma_{R/A}&\approx m_{\gamma}^2 \pm i\hat\omega m_{\gamma, 1}^2+\mathcal{O}(\hat\omega^2) ,\\ 
     \delta_{R/A}  &\approx  \frac{\hat{\omega}}{\tilde{n}}\left[\pm i m_{\delta}^2 +\hat\omega m_{\delta,1}^2\right]+\mathcal{O}(\hat\omega^3), \\ 
       \rho_{R/A}  &\approx  \frac{\hat{\omega}}{\tilde{n}}\left[\pm im_{\rho}^2 + \hat\omega m_{\rho,1}^2\right]+\mathcal{O}(\hat\omega^3).
    \end{align}
    \end{subequations}
In the small $\omega$ limit, the retarded (advanced) master integral $\mathcal{I}^{[n]}_{R/A}(\xi,\hat\omega,t,\hat{\nu})$ can be expanded as
\bea
\mathcal{I}^{[n]}_{R/A}(\xi,\hat{\omega},t,\hat{\nu}) \approx \Tilde{\mathcal{I}}^{[n]}(\xi,t,\hat{\nu}) \pm \hat\omega\  \Tilde{\mathcal{I}}^{[n]}_1(\xi,t,\hat{\nu})+\mathcal{O}(\hat\omega^2) ,
\eea
where
    \bea
\Tilde{\mathcal{I}}^{[n]}(\xi,t,\hat{\nu})&=&-\int_{0}^{1} dy \, \, \frac{2}{(1+ \xi y^2)^2} f^{[n]}(y,t,\hat{\nu}),\nn\\
\Tilde{\mathcal{I}}^{[n]}_1(\xi,t,\hat{\nu})&= &\int_{-1}^{1} dy \, \, \frac{y^n}{(1+ \xi y^2)^2} \frac{( t y-i\hat{\nu})\text{sign}(-t y)}{\left[(t y - i \hat{\nu})^2- (1-t^2) (1-y^2)\right]^{3/2}},\\
\Tilde{\mathcal{J}}(\xi,t,\hat{\nu})&=&\int_{-1}^{1} dy \, \, \frac{(\hat\nu+i t y)(i\hat\nu+t y\xi)}{(1+ \xi y^2)^2} \frac{\text{sign}(-t y)}{\left[(t y - i \hat{\nu})^2- (1-t^2) (1-y^2)\right]^{3/2}}.
\eea
Here,
\be
f^{[n]}(y, t, \hat{\nu})=\left\{\begin{array}{ll}
 \frac{\sqrt{s(y, t, \hat{\nu})+y^2-\hat{\nu}^2+t^2-1}}{\sqrt{2} s(y, t, \hat{\nu})}y^n, & n \text { is odd } \\
i \frac{\sqrt{2} \hat{\nu} t / s(y, t, \hat{\nu})}{\sqrt{s(y, t, \hat{\nu})+y^2-\hat{\nu}^2+t^2-1}}y^{n+1}, & n \text { is even }
\end{array},\right.
\ee
where $s(y, t, \hat{\nu}) \equiv \sqrt{\left(y^2-\hat{\nu}^2+t^2-1\right)^2+4 y^2 \hat{\nu}^2 t^2}$. So the mass scales become~\cite{Zhao:2023mrz}
 \begin{subequations}
\begin{align}
m_{\alpha}^2&=-\frac{\xi t \, m_D^2}{2(1-t^2)}\Big[tf_1(\xi)-(1-t^2+\hat{\nu}^2)\Tilde{\mathcal{I}}^{[1]}(\xi,t,\nu)-2 i \hat{\nu}t \Tilde{\mathcal{I}}^{[2]}(\xi,t,\hat{\nu})+\Tilde{\mathcal{I}}^{[3]}(\xi,t,\hat{\nu})\Big], \\
m_{\beta}^2&=\frac{m_D^2}{2\mathcal{W}(0,\hat{\nu})}\Big[f_0(\xi)-i \hat{\nu}\Tilde{\mathcal{I}}^{[0]}(\xi,t,\hat{\nu})-\xi t \Tilde{\mathcal{I}}^{[1]}(\xi,t,\hat{\nu})\Big], \\
            m_{\gamma}^2&=\frac{\xi \, m_D^2}{2(1-t^2)}\Big[(1+t^2)f_1(\xi)-t(1-t^2+2\hat{\nu}^2)\Tilde{\mathcal{I}}^{[1]}(\xi,t,\hat{\nu})-i\hat{\nu}(1+3t^2)\Tilde{\mathcal{I}}^{[2]}(\xi,t,\hat{\nu})
            +2t\Tilde{\mathcal{I}}^{[3]}(\xi,t,\hat{\nu})\Big],\\
            m_{\delta}^2&=\frac{m_D^2}{2\sqrt{1-t^2}}\Big[\hat{\nu}t f_0(\xi)-i\hat{\nu}^2 t\Tilde{\mathcal{I}}^{[0]}(\xi,t,\hat{\nu})+\hat{\nu}(1-\xi t^2)\Tilde{\mathcal{I}}^{[1]}(\xi,t,\hat{\nu})-i \xi t \Tilde{\mathcal{I}}^{[2]}(\xi,t,\hat{\nu})\Big], \\
            m_{\rho}^2&=-\frac{m_D^2}{2 \mathcal{W}(0,\hat{\nu})\sqrt{1-t^2}}\Big[\xi \hat{\nu}\Tilde{\mathcal{I}}^{[1]}(\xi,t,\hat{\nu})+i \xi t \Tilde{\mathcal{I}}^{[2]}(\xi,t,\hat{\nu})\Big],
        \end{align}
\end{subequations}
 %
  where the factor $\mathcal{W}(0,\hat{\nu})=(1 - \hat{\nu} \,\arccot\hat{\nu})$ 
  is real-valued, and hence, all mass scales are real-valued. Additionally, the first-order correction of the mass scales in the static limit is obtained as
 \begin{subequations}
 \begin{align}
  m_{\alpha,1}^2
  &=\frac{m_D^2}{2(1-t^2)}\left[\hat\nu f_0(\xi)+i(t^2-1-\hat\nu^2)\Tilde{\mathcal{I}}^{[0]}(\xi,t,\hat{\nu})+\hat{\nu}t(2-\xi) \Tilde{\mathcal{I}}^{[1]}(\xi,t,\hat{\nu})+i(1-t^2\xi)\Tilde{\mathcal{I}}^{[2]}(\xi,t,\hat{\nu})\right],\\
    m_{\beta,1}^2&= \frac{1}{ \mathcal{W}\left(0,\hat\nu\right)}\Bigg[
\frac{\hat\nu }{1+\hat\nu ^2}m_\beta^2 
+\frac{m_D^2}{2}\left\{\Tilde{\mathcal{J}}(\xi,t,\hat{\nu})+i\, \Tilde{\mathcal{I}}^{[0]}(\xi,t,\hat{\nu})\right\}\Bigg],\\
 m_{\gamma,1}^2 &=\frac{m_D^2}{2(1-t^2)} \Big[  -\hat{\nu}(1+t^2)f_0(\xi) + i(1-t^2+\hat{\nu}^2(1+t^2))\tilde{\mathcal{I}}^{[0]} - \hat{\nu} t (4-3\xi-\xi t^2) \tilde{\mathcal{I}}^{[1]} \\
& +i\xi t (1-t^2) \tilde{\mathcal{I}}_1^{[1]} -2 i(1-2\xi t^2 )\tilde{\mathcal{I}}^{[2]} - \hat{\nu} \xi (1+3t^2) \tilde{\mathcal{I}}_1^{[2]} - 2i\xi t \tilde{\mathcal{I}}_1^{[3]} \Big],\\
   m_{\delta,1}^2&= \frac{m_D^2}{2(1-t^2)} \Big[ t f_0 - 2 i \hat{\nu} t \tilde{\mathcal{I}}^{[0]} + \hat{\nu}^2 t \tilde{\mathcal{I}}_1^{[0]} + (1-\xi t^2) \tilde{\mathcal{I}}^{[1]} + i\hat{\nu} (1-\xi t^2) \tilde{\mathcal{I}}_1^{[1]} + \xi t \tilde{\mathcal{I}}_1^{[2]} \Big],\\
  m_{\rho,1}^2&=\frac{1}{2\mathcal{W}(0,\hat\nu)(1-t^2)} \left[ t f_0(\xi) -i \hat{\nu} t \tilde{\mathcal{I}}^{[0]} + (1-\xi t^2) \tilde{\mathcal{I}}^{[1]} - i\hat{\nu} \xi \tilde{\mathcal{I}}_1^{[1]} +  \xi t \tilde{\mathcal{I}}_1^{[2]}  + \frac{\hat\nu \left(\hat\nu\xi \tilde{\mathcal{I}}^{[1]}+it\xi \tilde{\mathcal{I}}^{[2]}\right)}{\mathcal{W}(0,\hat\nu)(1+\hat\nu^2)} \right]  .
     \end{align}
\end{subequations}
It is useful to note that the master integrals can be evaluated analytically in the collisionless limit, and the mass scales at $\nu=0$ become
 \begin{subequations}
\bea
\left.m_{\alpha}^2\right|_{\nu=0}
&=&-\frac{m_D^2t^2}{2 (1-t^2)}\left[f_0(\xi)-\frac{1}{ 1+\xi(1- t^2)}f_0\left(\frac{\xi t^2}{ 1+\xi(1- t^2)}\right) \right], \\
\left.m_{\beta}^2\right|_{\nu=0}
&=&\frac{m_D^2}{2} \left[f_0(\xi)+\frac{\xi t^2}{\left( 1+\xi(1- t^2)\right)^2}f_0\left(\frac{\xi t^2}{ 1+\xi(1- t^2)}\right)\right], \\
\left.m_{\gamma}^2\right|_{\nu=0}
&=&-\frac{m_D^2}{2(1-t^2)}\left[\frac{2(1-t^2)}{1+\xi}-(1+t^2)f_0(\xi)+\left(3-\frac{1}{1+\xi(1-t^2)}\right)\frac{t^2}{1+\xi(1-t^2)}f_0\left(\frac{\xi t^2}{1+\xi(1-t^2)}\right)\right],\\
\left.m_{\delta}^2\right|_{\nu=0}&=&\left.m_{\rho}^2\right|_{\nu=0}
= -\frac{\pi}{4}\frac{ m_D^2\xi t\sqrt{1-t^2}}{\left[1+\xi(1-t^2)\right]^{3/2}}.
\label{mass_scales_nu0}
\eea
 \end{subequations}
 and
 \begin{subequations}
\bea
\left.m_{\alpha,1}^2\right|_{\nu=0}&=&-\frac{m_D^2}{2}\frac{(1+\xi)}{2\left(1+\xi(1-t^2)\right)^{3/2}},\\
\left.m_{\beta,1}^2\right|_{\nu=0}&=&-\frac{m_D^2}{2}\frac{2+(3-t^2)\xi+(1-t^2)(1-2t^2)\xi^2}{2\left[1+\xi(1-t^2)\right]^{5/2}},\\
\left.m_{\gamma,1}^2\right|_{\nu=0}&=&\frac{\pi m_D^2}{2}\frac{\xi(1-t^2)(1+\xi(1+2t^2))}{2\left(1+\xi(1-t^2)\right)^{5/2}},\\
\left.m_{\delta,1}^2\right|_{\nu=0}&=&\left.m_{\rho,1}^2\right|_{\nu=0}\nn\\
&=&\frac{m_D^2}{2(1-t^2)}\left[\frac{3t(1-t^2)\xi}{[1+\xi(1-t^2)]^2}+\frac{t}{\sqrt{\xi}}\arctan\sqrt{\xi}\right.\nn\\
&&\left.\hspace{1cm}-\frac{1+\xi(1-t^2)-3\xi^2t^2(1-t^2)}{\sqrt{\xi}(1+\xi(1-t^2))^{5/2}}\arctan\frac{t\sqrt{\xi}}{\sqrt{1+\xi(1-t^2)}}\right]
\eea
\end{subequations}

\section{Dielectric Permittivity and Fluctuation Function}
\label{sec:eps}
In this section, we deduce the complex color dielectric response in a collisional anisotropic quark-gluon plasma medium, which is later used to determine the in-medium heavy-quark potential. In the temporal axial gauge, the inverse permittivity is computed from the spatial components of the gluon propagator in the static limit as:
\begin{equation}
\epsilon^{-1}(p) = -\lim_{\omega\to 0}\omega^2 \frac{p^i p^j}{p^2} \Delta_{ij}(\omega,p).
\label{eps_def}
\end{equation}
The real part of this response, $\text{Re}\,\epsilon^{-1}$, describes the causal screening of the potential and is obtained by taking the retarded propagator $\Delta = \Delta^R$. In the isotropic case, the imaginary part is directly related to medium fluctuations through the fluctuation-dissipation theorem. However, in an anisotropic medium, this relationship is more complex, which motivates the definition of the fluctuation function $\Phi(p, \xi, \nu)$ directly from the Feynman propagator $\Delta_F$ to capture the statistical field correlations.

\subsection{The real part of the inverse dielectric constant}
The real part of the inverse dielectric constant is obtained from the longitudinal projection of the retarded structure functions as:
\begin{equation}
\text{Re}\,\epsilon^{-1}(p) = -\lim_{\omega\to 0} \omega^2 \frac{p^i p^j}{p^2} \text{Re}\Delta^R_{ij}(\omega,p) = -\lim_{\omega\to 0}\omega^2 \beta'_R = -\lim_{\omega\to 0}\omega^2 \frac{(p^2-\omega^2+\alpha_R+\gamma_R)}{(\omega^2-\Omega_{+}^2)(\omega^2-\Omega_{-}^2)}.
\end{equation}
In the static limit, the real part of the dielectric permittivity can be written in terms of the mass scales:
\begin{equation}
\text{Re}\,\epsilon^{-1}(p) = \frac{p^2(p^2+m_{\alpha}^2+m_{\gamma}^2)}{p^4+p^2 \left(m_\alpha^2+m_\beta^2+m_\gamma^2\right)+m_\alpha^2 m_\beta^2+m_\beta^2 m_\gamma^2- m_\delta^2 m_\rho ^2} = \frac{p^2(p^2+m_{\alpha}^2+m_{\gamma}^2)}{(p^2+m_{+}^2)(p^2+m_{-}^2)},
\label{Re_eps_final}
\end{equation}   
where the effective mass scales $m_{\pm}$ are defined by:
\begin{equation}
2m^2_{\pm}=\left(m_{\alpha}^2+m_{\beta}^2+m_{\gamma}^2\right) \pm\sqrt{\left(m_{\alpha}^2+m_{\beta}^2+m_{\gamma}^2\right)^2-4\left\{m_{\beta}^2(m_{\alpha}^2+m_{\gamma}^2)-m_{\delta}^2m_\rho^2\right\}}.
\end{equation}
In the collisionless limit, the form of $\text{Re}\,\epsilon^{-1}(p)$ remains the same with $m_\delta = m_\rho$. In the isotropic collisional case where $m_\alpha = m_\gamma = m_\delta = m_\rho = 0$ and $m_\beta = m_D$, we obtain
\begin{equation}
\lim_{\xi\to 0}\text{Re}\,\epsilon^{-1}(p,\nu) = \frac{p^2}{p^2+m_D^2}.
\label{Re_eps_final_xi0}
\end{equation}

\subsection{The fluctuation function from the Feynman propagator}
The fluctuation function $\Phi(p, \xi, \nu)$ is defined from the static
limit of the longitudinal component of the Feynman propagator in the
temporal axial gauge:
\begin{equation}
\Phi(p, \xi, \nu) \equiv \frac{p^2}{2} \text{Im} \left[
\lim_{\omega \to 0} \omega^2
\frac{p^i p^j}{p^2}
\Delta_{ij}^F(\omega, p, \xi, \nu)
\right].
\label{Phi_def}
\end{equation}
It is important to emphasize the physical justification for defining the fluctuation function $\Phi(p,\xi,\nu)$ directly through the Feynman 
propagator. While the fluctuation-dissipation theorem (FDT) relating the symmetric and retarded/advanced propagators remains robustly valid for an 
isotropic collisional plasma ($\xi=0$), it breaks down once momentum anisotropy ($\xi \neq 0$) is introduced. Because the system is inherently out of equilibrium due to this directional anisotropy, 
evaluating the imaginary component of the heavy-quark potential requires a direct, first-principles calculation of the out-of-equilibrium Feynman 
(symmetric) propagator in the real-time formalism. We achieve this by employing the stochastic transport framework detailed in Appendix~\ref{struc_fun_F}, thereby completely bypassing the limitations of the standard equilibrium spectral function approach in the anisotropic regime.

This definition in Eq.~\eqref{Phi_def} is written in TAG, where the relevant quantity is the longitudinal projection of the spatial Feynman propagator, $\hat p^i\Delta_F^{ij}\hat p^j$.
Now, the one-loop effective Feynman propagator $\Delta_F^{ij}$ is given by
\begin{equation}
\begin{aligned}
\Delta_F^{ij}(P)
=&\left(1+2 f_B(\omega)\right)\mathrm{sgn}(\omega)
\left[\Delta_R^{ij}(P)-\Delta_A^{ij}(P)\right]
\\
&+\Delta_R^{ik}(P)
\left\{
\Pi_F^{kl}(P)
-\left(1+2 f_B(\omega)\right)\mathrm{sgn}(\omega)
\left[\Pi_R^{kl}(P)-\Pi_A^{kl}(P)\right]
\right\}
\Delta_A^{lj}(P).
\end{aligned}
\label{D_F_def}
\end{equation}
As a vital analytical check of our formulation, we note that the full Feynman propagator in Eq.~\eqref{D_F_def} and its corresponding form factors smoothly 
reproduce well-established limiting cases in the literature:
\begin{enumerate}
    \item In the collisionless anisotropic limit ($\xi \neq 0, \nu = 0$), the second term simplifies appropriately, and our formulation perfectly 
    reduces to the real-time anisotropic propagator results derived in 
   Ref.~\cite{Nopoush:2017zbu}.
    
    \item In the isotropic collisional limit ($\xi = 0, \nu \neq 0$), the second term vanishes identically, and our expression cleanly reduces to the collisional Feynman propagator obtained via the FDT from the retarded response calculated in Ref.~\cite{Carrington:2003je}.
\end{enumerate}
This successful cross-limital benchmarking reinforces the structural 
reliability and physical consistency of our generalized out-of-equilibrium 
framework across the entire parameter space.

We denote the longitudinal projection $\hat{p}^i \hat{p}^j M_{ij}$ of any tensor $M$ with the subscript $long$. Projecting Eq.~\eqref{D_F_def} onto the longitudinal direction and multiplying by $\omega^2$, the first term in the static limit becomes:
\bea
&&\lim_{\omega \to 0}\left[1+2 f_B(\omega)\right] \text{sgn}(\omega)\omega^2\left(\beta_R^{\prime}-\beta_A^{\prime}\right) \nn\\
&=&\frac{4 i p \lambda}{\sqrt{1+\xi(\bm{v} \cdot \bm{n})^2}} \frac{\left(m_{\alpha,1}^2+m_{\gamma,1}^2\right) m_\delta^2 m_\rho^2+\left(p^2+m_\alpha^2+m_\gamma^2\right)\left\{m_{\beta,1}^2\left(p^2+m_\alpha^2+m_\gamma^2\right)+(m_\delta^2 m_{\rho,1}^2+m_\rho^2 m_{\delta,1}^2)\right\}}{\left[\left(p^2+m_\beta^2\right)\left(p^2+m_\alpha^2+m_\gamma^2\right)-m_\delta^2 m_\rho^2\right]^2}.
\eea
Similarly, evaluating the last term of Eq.~\eqref{D_F_def} in the static limit yields:
\bea
&&\lim_{\omega \to 0}\left(1+2 f_B(\omega)\right) \text{sgn}(\omega)\omega^2\left[\Delta_R\left(\Pi_R-\Pi_A\right) \Delta_A\right]_{\rm long} \nn\\
&=&\frac{4 i p \lambda}{\sqrt{1+\xi(\bm{v} \cdot \bm{n})^2}} \frac{\left(m_{\alpha,1}^2+m_{\gamma,1}^2\right) m_\delta^2 m_\rho^2+\left(p^2+m_\alpha^2+m_\gamma^2\right)\left\{m_{\beta,1}^2\left(p^2+m_\alpha^2+m_\gamma^2\right)+(m_\delta^2 m_{\rho,1}^2+m_\rho^2 m_{\delta,1}^2)\right\}}{\left[\left(p^2+m_\beta^2\right)\left(p^2+m_\alpha^2+m_\gamma^2\right)-m_\delta^2 m_\rho^2\right]^2}.
\eea
Comparing these results shows that the first and last terms cancel each other in the static limit. The remaining contribution of $\omega^2\hat p^i\Delta_F^{ij}\hat p^j$ in the static limit can be obtained using the Eq.~\eqref{RFA_appen_B} of Appendix~\ref{app:tensor_relation} as
\begin{equation}
\lim_{\omega \to 0}\omega^2
\left[\Delta_R \Pi_F \Delta_A\right]_{\rm long}
=
\omega^2 \beta'_R \beta_F \beta'_A
+
\omega^2 \rho'_R (\alpha_F+\gamma_F)\delta'_A
\frac{p_\perp^2}{p^2}.
\label{eq:long_DRPFDA}
\end{equation}
The Feynman structure functions $\alpha_F, \beta_F,$ and $\gamma_F$ are discussed in Appendix~\ref{struc_fun_F}. In the static limit, we have:
\bea
\lim _{\omega \to 0}  \beta_F &=& -\frac{4 i \lambda m_D^2 \omega^2}{p^3} \frac{\langle 1 \rangle_\xi \langle \frac{\hat{\nu}}{(\bm{v}\cdot\hat{\bm{p}})^2+\hat{\nu}^2} \rangle_\xi}{\langle \frac{(\bm{v}\cdot\hat{\bm{p}})^2}{(\bm{v}\cdot\hat{\bm{p}})^2+\hat{\nu}^2} \rangle_\xi}, \\
 \lim_{\omega \to 0} (\alpha_F + \gamma_F) &=& -\frac{4i\lambda m_D^2}{p\,\tilde{n}^2} \left\langle \frac{\hat\nu(\bm{v}\cdot\tilde{\bm{n}})^2}{(\bm{v}\cdot\hat{\bm{p}})^2 + \hat\nu^2} \right\rangle_\xi.
\eea
Collecting these contributions, the fluctuation function is evaluated in two distinct regimes:

\subsubsection{Collisionless limit:}
When $\nu = 0$, the fluctuation function is expressed in terms of complete elliptic integrals:
\begin{equation}
\Phi(p, \xi, \nu=0) = \frac{2 \lambda m_D^2 p}{\varsigma \left[\left(p^2+m_\beta^2\right)\left(p^2+m_\alpha^2+m_\gamma^2\right)-m_\delta^2 m_\rho^2\right]^2} \left[\frac{m_\delta^2 m_\rho^2-\varsigma\left(p^2+m_\alpha^2+m_\gamma^2\right)^2}{1+\varsigma} E(-\varsigma)-m_\delta^2 m_\rho^2 K(-\varsigma)\right].
\end{equation}
Here, the variable $\varsigma = \xi(1-t^2)$ represents the effective anisotropy strength projected onto the transverse plane, and $t = \cos\theta$ is the cosine of the propagation angle.
\subsubsection{Collisional case:}
For a finite collision rate $\nu \neq 0$, the fluctuation function takes the form:
\bea
\Phi(p, \xi, \nu) &=& -\frac{2 \lambda m_D^2 p}{\left[\left(p^2+m_\beta^2\right)\left(p^2+m_\alpha^2+m_\gamma^2\right)-m_\delta^2 m_\rho^2\right]^2} \nn\\
&\times&\left[(p^2+m_\alpha^2+m_\gamma^2)^2 \frac{\left\langle 1 \right\rangle_\xi \left\langle \frac{\hat{\nu}}{(\bm{v}\cdot\hat{\bm{p}})^2+\hat{\nu}^2} \right\rangle_\xi}{\left\langle \frac{(\bm{v}\cdot\hat{\bm{p}})^2}{(\bm{v}\cdot\hat{\bm{p}})^2+\hat{\nu}^2} \right\rangle_\xi} + \frac{p^2 m_\delta^2 m_\rho^2}{p_\perp^2} \left\langle \frac{\hat\nu(\bm{v}\cdot\tilde{\bm{n}})^2}{(\bm{v}\cdot\hat{\bm{p}})^2 + \hat\nu^2} \right\rangle_\xi \right].
\label{Phi_final}
\eea

\subsection{Response and fluctuation in the isotropic limit}
In an isotropic plasma with collisions where $\xi=0$ and $\nu \neq 0$, the mass scales $m_\alpha, m_\gamma, m_\delta,$ and $m_\rho$ vanish. The real part of the inverse permittivity reduces to
\begin{equation}
\text{Re}\,\epsilon^{-1}(p, \xi=0, \nu) = \frac{p^2}{p^2+m_D^2}.
\label{Re_eps_nu}
\end{equation}
Similarly, the fluctuation function is obtained from Eq.~\eqref{Phi_final} as
\begin{equation}
\Phi(p, \xi=0, \nu) = -\frac{2T p m_D^2}{(p^2+m_D^2)^2} \frac{\arccot(\hat\nu)}{1-\hat\nu\arccot(\hat\nu)}.
\end{equation}
This expression can be verified through the fluctuation-dissipation theorem. In the isotropic collisional plasma, the retarded structure function $\beta_R$ is given by:
\begin{equation}
\beta_R(\omega,p,\xi=0, \nu)=-m_D^2\frac{\omega^2}{p^2}\left[1-\frac{\omega+i\nu}{2p}\ln\frac{\omega+p+i\nu}{\omega-p+i\nu}\right]\left[1-\frac{i\nu}{2p}\ln\frac{\omega+p+i\nu}{\omega-p+i\nu}\right]^{-1}.
\label{betaRxi0}
\end{equation}
In the small frequency limit, this structure function expands as
\begin{equation}
\beta_R(\omega, p, \xi=0, \nu) \approx -\hat{\omega}^2m_D^2\left[1+i\hat{\omega}\frac{\arccot(\hat{\nu})}{1-\hat{\nu}\arccot(\hat{\nu})}\right] + \mathcal{O}(\omega^4).
\end{equation}
The inverse permittivity then behaves as
\begin{equation}
\epsilon^{-1}(p, \xi=0, \nu) \approx \frac{p^2}{p^2+m_D^2} -i\omega \frac{ p m_D^2}{(p^2+m_D^2)^2} \frac{\arccot(\hat\nu)}{1-\hat\nu\arccot(\hat\nu)}.
\end{equation}
Applying the fluctuation dissipation theorem in the static limit, we relate the fluctuation function to the imaginary part of the permittivity as:
\begin{equation}
\Phi = \lim_{\omega \to 0} \frac{2T}{\omega} \text{Im}[\epsilon^{-1}].
\end{equation}
This recovers the same result and validates our derivation of the fluctuation function using the Feynman propagator.

\section{In-medium heavy quark-antiquark potential}
	\label{sec:pot}
In this section, we investigate the effect of a collisional anisotropic medium on the in-medium heavy-quark potential. We compute the in-medium heavy quark-antiquark potential by correcting the Cornell potential in momentum space using the dielectric permittivity (Eq.~\eqref{eps_def}), which accounts for the effects of the collisional anisotropic thermal medium~\cite{Agotiya:2008ie,Thakur:2013nia, Thakur:2016cki} 
\bea
 V({\bm r,\xi, \nu})=\int \frac{d^3{\bm p}}{(2\pi)^{3/2}} ~(e^{i{\bm {p\cdot r}}}-1)~\frac{ V_{\text{C}}(p)}{\epsilon(p, \xi, \nu)},\label{med_pot}
\eea
where $ V_{\rm C}(p)=-\sqrt{(2/\pi)}(\alpha/p^2 +2 \sigma/p^4 )$~\cite{Agotiya:2008ie}
is the Fourier transform of the vacuum Cornell potential $ V_{\rm C}(r)=-\alpha/r +\sigma \,r $. 
Here $\alpha=C_F\,\alpha_s $ with $C_F=4/3$ is the quadratic Casimir of the fundamental representation and $\alpha_s$ is the strong coupling constant~\cite{Haque:2014rua}
 \begin{equation}
     \alpha_s(\Lambda^{2})=\frac{12\pi}{(11N_{c}-2 N_{f})\ln\left(\frac{\Lambda^{2}}{\Lambda_{\overline{\rm MS}}^{2}}\right)},
		\label{alpha}
     \end{equation}
In Eq.~\eqref{alpha}, $N_c=3 $ and $\Lambda_{\overline{\rm MS}}=0.176 $ GeV for $ N_f=3 $ and $\Lambda=2\pi \lambda$. 
Here the string tension, $\sigma$, is taken as $(0.44)^2 $ GeV$^2$.
After substituting Eq.~\eqref{Re_eps_final} into Eq.~\eqref{med_pot}, we obtain the medium-modified complex heavy-quark potential in an anisotropic medium, whose real part reads
 \bea
\Re V(\bm{r},\xi,\nu)&=& \int \frac{d^3{\bm p}}{(2\pi)^{3/2}} ~(e^{i{\bm {p\cdot r}}}-1)~ V_{\text{C}}(p){\rm Re}\,\epsilon^{-1}(p,\xi, \nu)\nn\\
&=& -\frac{1}{2\pi^2}\int d^3{\bm p}\,~(e^{i{\bm {p\cdot r}}}-1)\frac{(p^2+m_{\alpha}^2+m_{\gamma}^2)}{(p^2+m_{+}^2)(p^2+m_{-}^2)} \left(\alpha+\frac{2\sigma}{p^2}\right).
\label{ReV_def}
\eea
For a general orientation of $\bm{r}$ relative to the anisotropy direction $\bm{\hat n}$, we will have,
$\bm{p}\cdot\bm{r}=rp [\sin \theta \sin \Theta \cos (\phi - \Phi)+ \cos \theta \cos \Theta]$, where $\theta(\Theta)$ and $\phi (\Phi)$ are the polar and the azimuthal angles in momentum (coordinate) space with the anisotropy direction, respectively, since we take $\bm{\hat{n}}=\hat{z}$. After integration over the azimuthal angle of $(\exp{(i \bm{p}\cdot r)}-1)$, Eq.~\eqref{ReV_def} becomes
\bea \label{real_pot}
\Re V(\bm{r},\xi,\nu,\lambda)&=&-\frac{1}{\pi}\int_{p=0}^\infty\int_{\theta=0}^{\pi} p^2\ dp\ \sin\theta d\theta  \left[e^{i p r \, \cos \theta \cos \Theta} J_0 (p r \sin \theta \sin \Theta)-1\right]\nn\\
&&\times\frac{(p^2+m_{\alpha}^2+m_{\gamma}^2)}{(p^2+m_{+}^2)(p^2+m_{-}^2)} \left(\alpha+\frac{2\sigma}{p^2}\right),\nn\\
&=&-\frac{2}{\pi}\int_{p=0}^\infty\int_{t=0}^{1} p^2\ dp\  dt  \left[e^{i p r \, t \cos \Theta} J_0 (p r \sqrt{1-t^2} \sin \Theta)-1\right]\nn\\
&&\times\frac{(p^2+m_{\alpha}^2+m_{\gamma}^2)}{(p^2+m_{+}^2)(p^2+m_{-}^2)} \left(\alpha+\frac{2\sigma}{p^2}\right),
\eea
where $J_0$ is the Bessel function of the first kind. 
The Coulomb part of Eq.~\eqref{real_pot} contains a $\lambda=0$ UV divergence, which we subtract, giving
\bea \label{real_pot_2}
\Re V(\bm{r},\xi,\nu,\lambda)&=&-\lim_{\Lambda_{\rm UV}\to\infty} \frac{2}{\pi}\int_{p=0}^{\Lambda_{\rm UV}}\int_{t=0}^{1} p^2\ dp\  dt  \left[e^{i p r \, t \cos \Theta} J_0 (p r \sqrt{1-t^2} \sin \Theta)-1\right]\nn\\
&&\times\frac{(p^2+m_{\alpha}^2+m_{\gamma}^2)}{(p^2+m_{+}^2)(p^2+m_{-}^2)} \left(\alpha+\frac{2\sigma}{p^2}\right)- \frac{2}{\pi} \int_{p=0}^{\Lambda_{\rm UV}}dp\,\alpha.
\eea
Here, $\Lambda_{\rm UV}$ is the UV cutoff scale and Eq.~\eqref{real_pot_2} is independent of the scale for a large value of $\Lambda_{\rm UV}$. Note that for a collisionless, isotropic plasma, the real part of the quarkonium potential is
\bea \label{real_pot_HTL}
\Re V_{\rm{HTL}}(r,T)&=&-\alpha m_D\left(1+ \frac{e^{-m_Dr}}{m_Dr}\right)+\frac{2 \sigma }{m_D}\left(1+\frac{e^{-m_Dr}-1}{m_Dr}\right).
\eea
     \begin{figure}[h]
		\begin{center}
			\includegraphics[scale=.7]{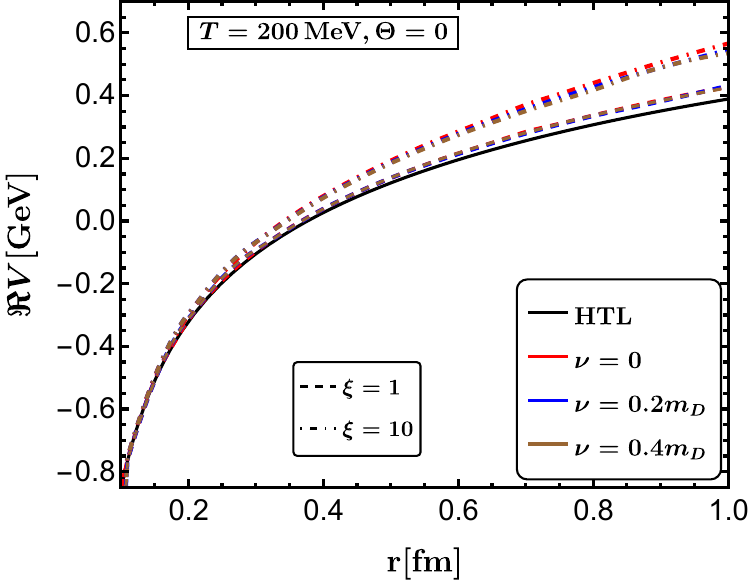}
            \includegraphics[scale=.7]{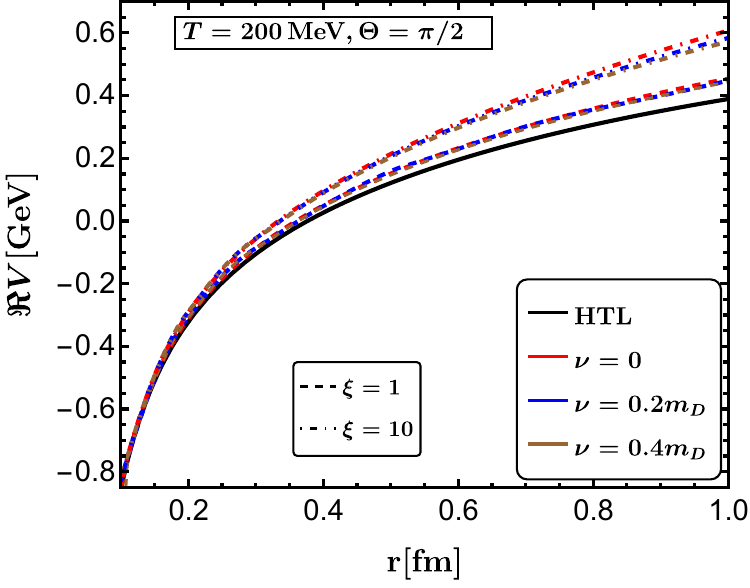}
			\caption{The real part of the potential as a function of separation distance $r$ for various values of $\nu$ and $\xi$ when $Q\bar{Q}$ is aligned parallel ($\Theta=0$) (left) and aligned transverse ($\Theta=\pi/2$) (right) to the direction of anisotropy at $T=0.2$ GeV. In both the figures, the dashed lines represent the $\xi=1$ case, whereas dot-dashed lines represent the $\xi=10$ case.}
			\label{fig_real1}
		\end{center}
	\end{figure}
In Fig.~\ref{fig_real1}, we plot the real part of the potential [Eq.~\eqref{real_pot_2}] as a function of the separation distance $( r) $ at temperature $\lambda=0.2$ GeV.
 The left panel illustrates the results for various values of the collision parameter, $\nu$, and anisotropy parameter, $\xi$, when the $Q\bar{Q}$ dipole axis is aligned parallel to the anisotropy axis, while the right panel shows the results when the dipole is aligned transverse to the anisotropy axis.
Here, the solid black line shows the real part of the potential for the isotropic collisionless plasma [Eq.~\eqref{real_pot_HTL}] when $\xi=0$ and $\nu=0$, while different colors denote the various values of the collision parameter, $\nu$. We find that the collisional parameter $\nu$ has only a nominal effect on the real part of the potential. Notably, the impact of $\nu$ on the real part of the potential remains nearly consistent for both parallel and transverse alignments. It is also worth mentioning here that for an isotropic plasma ($\xi=0$), there is no effect of the collision on the real part of the quarkonium potential, as can be understood from Eq.~\eqref{Re_eps_nu}.

\begin{figure}[h]
         \centering
         \includegraphics[width=0.5\linewidth]{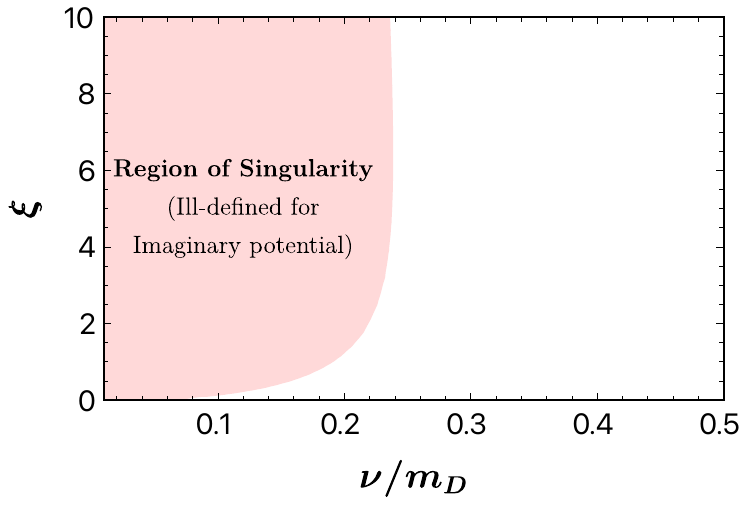}
         \caption{Region in the anisotropy-collision-parameter plane where the integral in Eq.~\eqref{im_pot} develops a singularity.}
         \label{fig:singularity}
     \end{figure}

The imaginary part of the potential is written as 
\bea\label{im_pot}
\Im \,V(\bm{r},\xi,\nu,T) &=& \int \frac{d^3 p}{(2 \pi)^{3/2}} (e^{i \bm{p}\cdot \bm{r}}-1) V_C(p) \text{Im}\,\epsilon^{-1}(p,\xi, \nu)\nn\\
&=& -\frac{1}{\pi} \int_{p=0}^{\infty} \, \int_{\theta=0}^{\pi} p^2 \, dp\, \sin\theta\,d\theta \, \left[e^{i p r \, \cos \theta \cos \Theta} J_0 (p r \sin \theta \sin \Theta)-1\right] \nn \\
&\times& \frac{1}{p^2} \left(\alpha + \frac{2 \sigma}{p^2}\right)  \text{Im}\,\epsilon^{-1}(p,\xi, \nu) \nn\\
&=&-\frac{2}{\pi} \int_{p=0}^{\infty} \, \int_{t=0}^{1} p^2 \, dp\, \, dt \, \left[\cos\left( p \, r \, t \cos \Theta\right) J_0 (p r \sqrt{1-t^2} \sin \Theta)-1\right]\frac{1}{p^2} \left(\alpha + \frac{2 \sigma}{p^2}\right) \nn \\
&&\hspace{-1cm}\times  \frac{-2  \lambda m_D^2p}{\left(p^2+m_+^2\right)^2\left(p^2+m_-^2\right)^2}\left[(p^2+m_\alpha^2+m_\gamma^2)^2\frac{\left\langle1\right\rangle_\xi\left\langle\frac{\hat{\nu}}{(\bm v\cdot\hat{\bm p})^2+\hat{\nu}^2}\right\rangle_\xi}{\left\langle\frac{(\bm{v\cdot\hat{p}})^2}{(\bm{v\cdot\hat{p}})^2+\hat{\nu}^2}\right\rangle_\xi}+ \frac{p^2m_\delta^2 m_\rho^2}{p_\perp^2}\left\langle \frac{\hat\nu(\bm{v}\cdot\bm{\tilde{n}})^2}{(\bm{v}\cdot\bm{\hat p})^2 + \hat\nu^2} \right\rangle_\xi \right],
\eea
where $t=\cos \theta$.  Note that for the isotropic case, the imaginary potential reduces to 
\bea\label{im_pot_nu}
\left.\text{Im} \,V(r,T)\right|_{\xi\to 0} &=& -2Tm_D^2\int \frac{p\ dp}{\left(p^2+m_D^2\right)^2}\left(\alpha+\frac{2\sigma}{p^2}\right)\left(1-\frac{\sin pr}{pr}\right)\frac{2\arccot\hat\nu}{\pi\left(1-\hat\nu\arccot\hat\nu\right)}.
\eea
Equation~\eqref{im_pot_nu} further reduces to the collisionless case as
\bea\label{im_pot_HTL}
\text{Im} \,V_\text{HTL}(r,T) &=& -2Tm_D^2\int \frac{p\ dp}{\left(p^2+m_D^2\right)^2}\left(\alpha+\frac{2\sigma}{p^2}\right)\left(1-\frac{\sin pr}{pr}\right).
\eea
In Eq.~\eqref{im_pot}, the integral has a pinch singularity. This singularity arises from the term $(p^2+m_{-}^2)^2$ in the denominator of the integrand, which makes the imaginary part of the potential ill-defined. The $(p^2+m_{-}^2)$ term also appears in the integral of real potential in Eq.~\eqref{ReV_def}, where it is a simple pole and can be integrated using the principal-value prescription. But for the imaginary potential, $(p^2+m_{-}^2)^2$ appears as a double pole; hence, it cannot be integrated out using the principal-value prescription. The issue of the pinch singularity was previously discussed in~\cite{Nopoush:2017zbu} in the collisionless case, that is, when $\nu=0$. In Fig.~\ref{fig:singularity}, we illustrate the region where this singularity occurs for the finite collisional effect, which certainly includes the $\nu=0$ axis as mentioned in~\cite{Nopoush:2017zbu} for the collisionless case. In this way, we identify the region of anisotropic collisional plasma, where the imaginary potential is calculable, and the Weibel instability is absent. In our numerical observation, we found that the region of singularity is extended approximately from $\nu=0$ up to $\nu \sim 0.24 m_D$ for anisotropy parameter $\xi \sim 10$. In this article, we show the imaginary part of the potential only for the region where the Weibel instability is absent, i.e., for the white colored region in Fig.~\ref{fig:singularity}. In Fig.~\ref{fig:singularity}, the right edge can be obtained by reaching closer to $p\to 0$ and $t\to 1$. As $p$-value increases and the $t$-value decreases, the singularities move toward $\nu \to 0,$ and ultimately all singularities accumulate on the $\nu=0$ axis.
 \begin{figure}[tbh]
		\begin{center}
			\includegraphics[scale=.68]{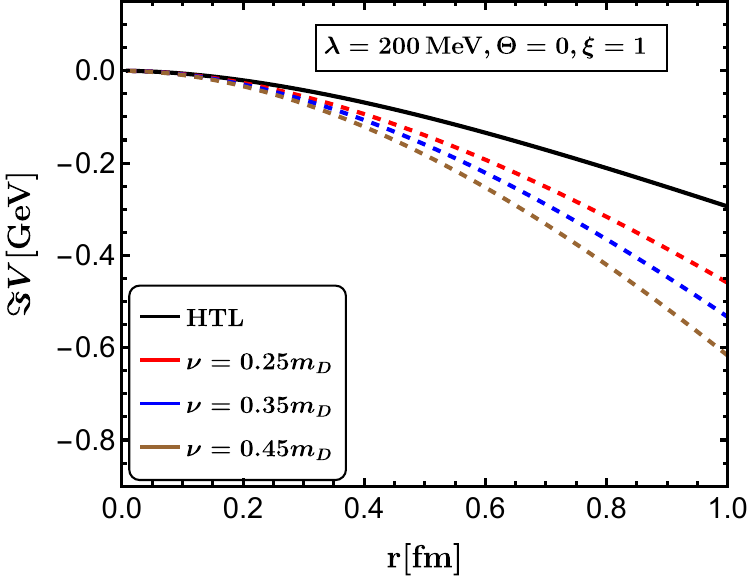}
            \includegraphics[scale=.68]{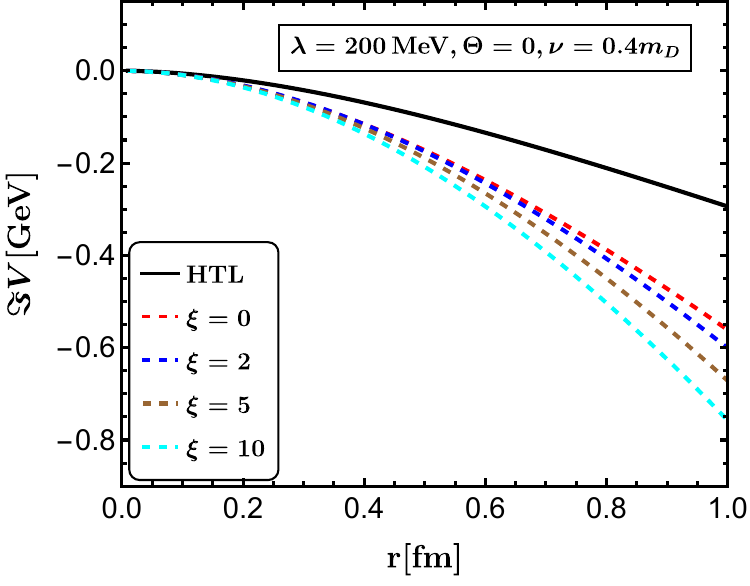}
			\caption{The imaginary part of the potential as a function of separation distance $r$ for different values of $\nu$ (left) and $\xi$ (right) at $\lambda=0.2$ GeV and $\Theta=0$.}
			\label{fig_im1}
		\end{center}
	\end{figure}
In Fig.~\ref{fig_im1}, we plot the imaginary part of the potential in Eq.~\eqref{im_pot} as a function of separation distance $r$ at temperature $T=0.2$ GeV when the dipole axis $ Q\bar{Q} $ is aligned parallel to the direction of anisotropy. The left panel shows the results for various values of $\nu$ while keeping $ \xi = 1$. We find that the imaginary part of the potential increases in magnitude as $\nu$ increases.  
The right panel presents the results for the different values of $\xi$. The dash-dotted lines correspond to $ \nu = 0.4 m_D $. 
In this case, the imaginary part of the potential 
increases in magnitude with the increase in $\xi$.
The solid black line shows the imaginary part of the potential for the isotropic collisionless plasma~\eqref{im_pot_HTL}.
\begin{figure}[tbh]
		\begin{center}
			\includegraphics[scale=.68]{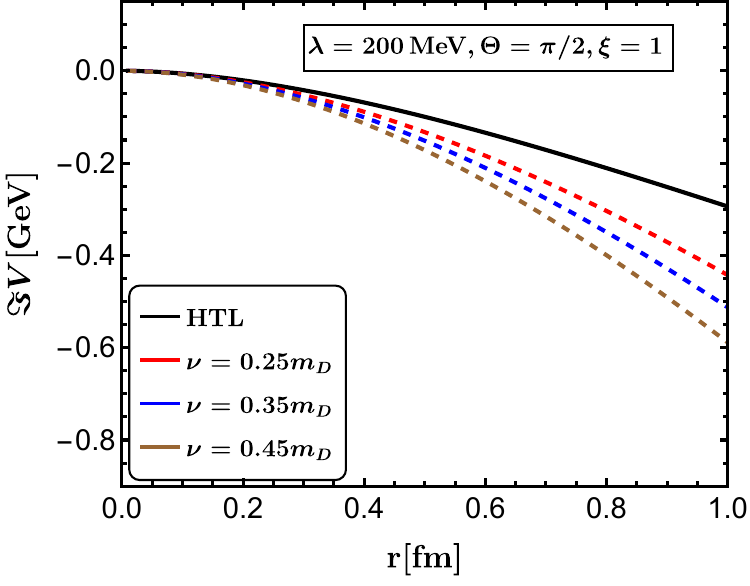}
            \includegraphics[scale=.68]{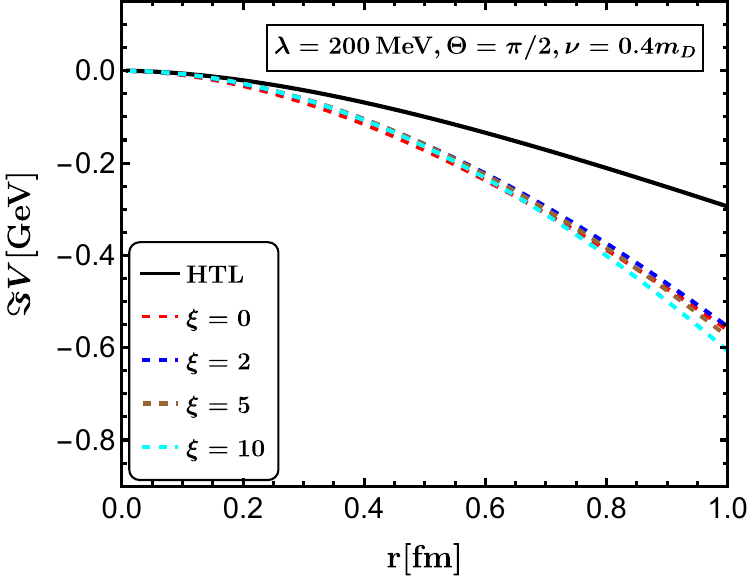}
			\caption{The imaginary part of the potential as a function of $r$ for different values of $\nu$ (left) and $\xi$ (right) at $\lambda=0.2$ GeV and $\Theta=\pi/2$.}
			\label{fig_im2}
		\end{center}
	\end{figure}
Figure~\ref{fig_im2} shows a similar plot of the imaginary part of the potential for the perpendicular orientation of the $Q\bar{Q}$ dipole axis relative to the direction of anisotropy. The effect of $\nu$ on the imaginary part of the potential is similar for both the parallel and perpendicular orientations. Notably, the impact of $\nu$ is more pronounced on the imaginary part of the potential compared to the real part, where its effect is minimal. 
 
	\section{Decay Width}
	\label{sec:dw}
In this section, we determine the decay width of quarkonium states based on the imaginary part of the potential obtained earlier. Treating this imaginary component as a perturbation to the vacuum potential allows us to estimate the thermal width, $\Gamma$, of the quarkonium states as
 \bea
\Gamma (\xi, \nu, \lambda)&=&- \int d^3\bm{r} |\psi(r,\lambda)|^2 \text{Im}V(\bm{r},\xi, \nu,\lambda)\nn \\
&=& - 2\pi \int_{r=0}^{\infty} dr\, r^2 \, \int_{\Theta=0}^{\pi} d\Theta \, \sin(\Theta) \, |\psi(r,\lambda)|^2 \, \text{Im}V(\bm{r}, \Theta , \xi, \nu ,\lambda) \nn \\
&=& - \frac{2}{a_0^3 } \int_{r=0}^{\infty} dr\, r^2 \, \int_{\Theta=0}^{\pi} d\Theta \, \sin(\Theta) \, e^{-2r/a_0(\lambda)} \, \text{Im}V(\bm{r}, \Theta, \xi,\nu, \lambda) \nn \\
&=& \frac{4}{\pi\, a_0^3 } \int_{r=0}^{\infty} dr\, r^2 \, \int_{\Theta=0}^{\pi} d\Theta \, \sin(\Theta) \, \text{e}^{-2r/a_0(\lambda)} \, \int_{p=0}^{\infty}\int_{t=0}^{1} \, dp\, \, dt \nn \\
&\times& \left[\cos\left( p \,r \, t \cos \Theta\right) J_0 (p r \,\sqrt{1-t^2} \sin \Theta)-1\right]\left(\alpha+ \frac{2\sigma}{p^2}\right) \Phi(p,\xi,\nu,t),
\label{dw_with_r}
 \eea
where we have used the Coulombic ground state wave function
\be
\psi(r,\lambda)=\frac{1}{\sqrt{\pi \,a_0(\lambda)^3}}\text{e}^{-r/a_0(\lambda)}, \quad \text{and} \quad a_0(\lambda)=\frac{2}{m_0 \alpha(\lambda)}
\ee
is the Bohr radius of the heavy quarkonium system, and $m_0$ is the heavy-quark mass. In the $\overline{\text{MS}}$ scheme, the charm and bottom quark masses are taken as $1.275$ GeV and $4.18$ GeV, respectively. The primary contribution to the imaginary potential for a deeply bound state involving heavy quarks is Coulombic in nature. This supports the use of Coulomb wave functions for calculating the decay width.

The $r$-integration in Eq.~\eqref{dw_with_r} can be computed using the integral
\bea
\int_0^\infty r^2 e^{-a r}J_0(cr)dr=\frac{2a^2-c^2}{\left(a^2+c^2\right)^{5/2}}.
\label{int_rel}
\eea
The integral relation in Eq.~\eqref{int_rel} can be used to perform the following integration as
\bea
\int_0^\infty r^2 e^{-a r} \cos(br)J_0(cr)dr=\frac{1}{2}\left[\frac{2(a+ib)^2-c^2}{\left((a+ib)^2+c^2\right)^{5/2}}+\frac{2(a-ib)^2-c^2}{\left((a-ib)^2+c^2\right)^{5/2}}\right].
\label{int_rel2}
\eea
In Eq.~\eqref{dw_with_r},
\begin{subequations}
    \bea
    a&=&\frac{2}{a_0},\\
     b&=& p  \, t \cos \Theta,\\
      c&=&p  \,\sqrt{1-t^2} \sin \Theta.
    \eea
\end{subequations}
So, Eq.~\eqref{dw_with_r} becomes 
\bea
\label{eq:DW}
\Gamma(\xi,\nu,\lambda)&=&\int\limits_{p=0}^\infty\int\limits_{t=0}^1\int\limits_{\Theta=0}^\pi \frac{4 \sin \Theta}{\pi a_0^3}\left[\frac{\left(\frac{2}{a_0}-i p t \cos\Theta\right)^2-\frac{1}{2} p^2 \left(1-t^2\right) \sin ^2\Theta }{\left\{\left(\frac{2}{a_0}-i p t \cos\Theta\right)^2+ p^2 \left(1-t^2\right) \sin ^2\Theta \right\}^{5/2}}\right.\nn\\
&&\hspace{1cm}+\left.\frac{\left(\frac{2}{a_0}+i p t \cos\Theta\right)^2-\frac{1}{2} p^2 \left(1-t^2\right) \sin ^2\Theta }{\left\{\left(\frac{2}{a_0}+i p t \cos\Theta\right)^2+ p^2 \left(1-t^2\right) \sin ^2\Theta \right\}^{5/2}}-\frac{a_0^3}{4}\right]\left(\alpha+\frac{2\sigma}{p^2}\right)\Phi(p,\xi,\nu,t)dp\,dt\,d\Theta\nonumber\\
&=&\frac{2}{\pi}\int\limits_{p=0}^\infty\int\limits_{t=0}^1\left[\frac{16}{\left(a_0^2p^2+4\right)^2}-1\right]\left(\alpha+\frac{2\sigma}{p^2}\right)\Phi(p,\xi,\nu,t)dp\,dt.
\eea
    \begin{figure}[h]
        \centering
        \includegraphics[width=0.47\linewidth]{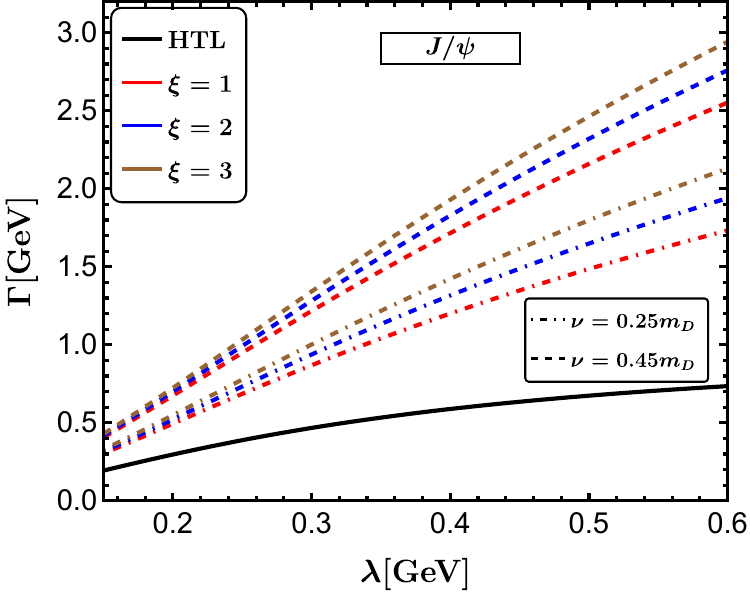}
\includegraphics[width=0.47\linewidth]{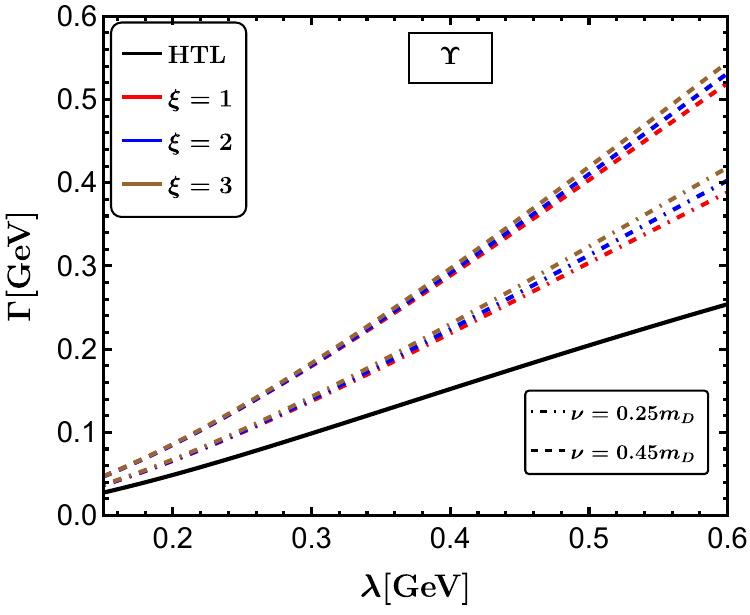}
        \caption{Temperature variation of decay width for the ground state of charmonium (left) and bottomonium (right).}
        \label{fig:decay_width}
    \end{figure}
For an isotropic, collisionless medium, the decay width of quarkonium can be obtained analytically as
\bea
\label{eq:DW_HTL}
\Gamma_{\rm{HTL}}&=& \frac{2}{\pi}\int\limits_{p=0}^\infty\int\limits_{t=0}^1\left[\frac{16}{\left(a_0^2p^2+4\right)^2}-1\right]\left(\alpha+\frac{2\sigma}{p^2}\right)\frac{- \pi T p m_D^2}{(p^2 + m_D^2)^2}\, dp\,dt\nonumber\\
&=& \frac{\alpha m_D^2 a_0^2 T}{\(a_0^2 m_D^2 - 
4\)}\left[1-\frac{8}{\(a_0^2 m_D^2 - 4\) }
 + \frac{64}{\(a_0^2 m_D^2 - 4\)^2}\log\frac{m_D a_0}{2}\right]\nonumber\\
 &&+\frac{4\sigma a_0^2 T}{\(a_0^2 m_D^2 - 4\)^2}\Bigg[4 + 
\frac{a_0^2 m_D^2\(a_0^2 m_D^2 - 12\)}{\(a_0^2 m_D^2 - 4\)}
\log\frac{m_D a_0}{2} \Bigg].
\eea
In Fig.~\ref{fig:decay_width}, we present the thermal decay width $\Gamma$ as a function of the temperaturelike scale $\lambda$ for the ground states of charmonium (left panel) and bottomonium (right panel), focusing on the parameter space where the imaginary potential is free from Weibel instabilities. The solid black line in both panels denotes the decay width calculated for an isotropic and collisionless medium, which serves as a baseline for our analysis. We observe that the decay width increases monotonically with both the temperature and the collisional parameter $\nu$, and the magnitude of the thermal width is further enhanced as the anisotropy parameter $\xi$ increases. This behavior is consistent with the characteristics of the imaginary part of the potential discussed previously, in which the interplay between collisions and anisotropy significantly modifies the thermal fluctuations within the medium. Quantitatively, the $J/\psi$ state exhibits a significantly larger decay width compared to the $\Upsilon$ state because the more tightly bound bottomonium system is more resilient to the screening and collisional effects of the plasma. These results suggest that charmonium dissociates more rapidly in a collisional, anisotropic environment, thereby supporting the experimental observation of sequential suppression in relativistic heavy-ion collisions.

\section{Summary and discussion}
\label{sec:summary}
In this work, we have investigated the complex heavy-quark potential within a quark-gluon plasma characterized by momentum-space anisotropy and finite collisional effects. We incorporated the anisotropic correction using the Romatschke-Strickland distribution function for thermal partons. The gluon self-energy was subsequently rederived by solving the linearized Boltzmann-Vlasov kinetic equations, including a Bhatnagar-Gross-Krook collision kernel. By computing the resummed gluon propagator, we determined the real and imaginary components of the dielectric permittivity in the static limit, which provided the necessary framework to derive the complex potential for heavy quarkonia.

Our results indicate that the momentum-space anisotropy, parametrized by $\xi$, reduces the overall screening effect in the medium. This anisotropy also introduces a directional dependence in the potential, which varies according to the angle $\Theta$ between the $Q\bar{Q}$ dipole axis and the anisotropy vector. While the anisotropy significantly modifies the screening behavior, we observed that the impact of collisions on the real part of the potential is nominal. This suggests that the binding energy of the quarkonium states is primarily sensitive to the anisotropy rather than the collisional rate.

In contrast, the imaginary component of the potential is significantly influenced by the collisional parameter $\nu$. We found that in an anisotropic medium, the imaginary potential can become ill-defined for certain ranges of the collision parameter due to the presence of a pinch singularity. However, within the well-defined regimes, the magnitude of the imaginary part increases with increasing collision frequency. This enhancement directly affects the thermal properties of the bound states.

Furthermore, we calculated the thermal decay width of the $J/\psi$ and of the $\Upsilon$ state. Our qualitative study, using a Coulombic wave function, demonstrates that the decay width increases with both the anisotropy and the collision rate among plasma constituents. These results imply that charmonium states undergo faster dissociation in a collisional plasma than in a collisionless one. 

To improve the quantitative accuracy of these results, future work could involve solving the Schr\"odinger equation with the full complex heavy-quark potential to obtain the modified wave functions and more precise decay widths. Additionally, further research is needed to explore more sophisticated nonequilibrium distribution functions and their consequences for the real and imaginary dielectric permittivity of the medium.

\acknowledgments{L.~T.~ would like to acknowledge the support of the National Research Foundation (NRF), funded by the Ministry of Science of Korea (Grant No. 2021R1F1A1061387). M.~D. and N.~H. acknowledge the support from DAE, Govt. of India.  }
	%

\appendix
\section{Calculations of structure functions}
\label{appen_A}
      We take anisotropy direction $\bm{n}= \hat{z}$ and assume $\bm{p}$ lies in the $(yz)$ plane, i.e., $\bm{p}=(0,p_y,p_z)$. In general, $p_y$ should be replaced by $p_{\perp}$:
\bea
\beta &=& \hat{p}^i \hat{p}^j m_D^2 \int \frac{d\Omega}{4\pi} v^i \frac{v^l + \xi (\bm{v}\cdot\bm{n})n^l}{(1+ \xi (\bm{v}\cdot\bm{n})^2)^2}\frac{\delta^{lj}(\hat{\omega}-\bm{\hat{k}}\cdot\bm{v})+\hat{k}^l v^j}{\hat{\omega}-\hat{\bm{k}}\cdot \bm{v}+i \hat{\nu}} \nn \\
&+& \hat{p}^i \hat{p}^j m_D^2 (i\hat{\nu})\int\frac{d\Omega'}{4\pi} \frac{v'^i}{\hat{\omega}-\hat{\bm{k}}\cdot\bm{v'}+i \hat{\nu}} \int \frac{d\Omega}{4\pi} \frac{v^l + \xi (\bm{v}\cdot\bm{n})n^l}{(1+ \xi (\bm{v}\cdot\bm{n})^2)^2} \frac{\delta^{lj}(\hat{\omega}-\bm{\hat{k}}\cdot\bm{v})+\hat{k}^l v^j}{\hat{\omega}-\hat{\bm{k}}\cdot \bm{v}+i \hat{\nu}} \, \mathcal{W}(\hat{\omega},\hat{\nu})^{-1} \nn \\
&=& m_D^2 \int \frac{d\Omega}{4\pi} (\bm{\hat{p}}\cdot\bm{v}) \frac{v^l+\xi \cos\theta \, n^l}{(1+\xi \cos^2 \theta)^2}\frac{\hat{p}^l(\hat{\omega}-\bm{\hat{p}}\cdot \bm{v})+\hat{p}^l(\bm{\hat{p}}\cdot\bm{v})}{\hat{\omega}-\bm{\hat{p}}\cdot\bm{v}+i\hat{\nu}} + m_D^2 \int \frac{d\Omega'}{4\pi} \frac{\bm{\hat{p}}\cdot \bm{v}' \, (i\hat{\nu})}{\hat{\omega}-\bm{\hat{p}}\cdot\bm{v}'+i\hat{\nu}} \nn \\
&\times& \int \frac{d\Omega}{4\pi} \frac{v^l+\xi \, \cos\theta \, n^l}{(1+\xi \cos^2\theta)^2} \frac{\hat{p}^l\, (\hat{\omega}-\bm{\hat{p}}\cdot \bm{v})+\hat{p}^l(\bm{\hat{p}}\cdot\bm{v})}{\hat{\omega}-\bm{\hat{p}}\cdot\bm{v}+i\hat{\nu}}\mathcal{W}(\hat{\omega},\hat{\nu})^{-1} \nn \\
&=& m_D^2 \int \frac{d\Omega}{4\pi} \hat{\omega} \left(\frac{\hat{\omega}+i\hat{\nu}}{\hat{\omega}+i\hat{\nu}-\bm{\hat{p}}\cdot\bm{v}}-1\right) \frac{\bm{\hat{p}}\cdot\bm{v}+\xi \, t \, \cos\theta}{(1+\xi \cos^2\theta)^2}+ (i\hat{\nu}) m_D^2 \int \frac{d\Omega'}{4\pi} \left(\frac{\hat{\omega}+i\hat{\nu}}{\hat{\omega}+i\hat{\nu}-\bm{\hat{p}}\cdot\bm{v}'}-1\right) \nn \\
&\times& \int \frac{d\Omega}{4\pi} \frac{\bm{\hat{p}}\cdot\bm{v}+\xi \, t\, \cos\theta}{(1+\xi \cos^2\theta)^2} \frac{\hat{\omega}}{\hat{\omega}+i\hat{\nu}-\bm{\hat{p}}\cdot\bm{v}} \mathcal{W}(\hat{\omega},\hat{\nu})^{-1} \nn \\
&=& \beta_1 +\beta_2.
\eea
Here,
\bea
\beta_1&=&m_D^2 \int \frac{d\Omega}{4\pi} \hat{\omega} \left(\frac{\hat{\omega}+i\hat{\nu}}{\hat{\omega}+i\hat{\nu}-\bm{\hat{p}}\cdot\bm{v}}-1\right) \frac{\bm{\hat{p}}\cdot\bm{v}+\xi \, t \, \cos\theta}{(1+\xi \cos^2\theta)^2} \nn \\
&=& \hat{\omega} m_D^2 \left[\int_{-1}^{1}\frac{dx}{2}\int_{0}^{2\pi} \frac{d\phi}{2\pi} \frac{1}{(1+\xi \, x^2)^2}\frac{z(z+\xi\,t\,x )}{z-\bm{\hat{p}}\cdot\bm{v}}- \int_{-1}^{1}\frac{dx}{2}\int_{0}^{2\pi} \frac{d\phi}{2\pi} \frac{z}{(1+\xi\,x^2)^2}\right] \nn \\
&=& \hat{\omega} m_D^2 \left[\int_{-1}^{1}\frac{dx}{2}\int_{0}^{2\pi} \left(\frac{z^2}{z-\bm{\hat{p}}\cdot\bm{v}}+ z\xi \, t \frac{x}{z-\bm{\hat{p}}\cdot\bm{v}}\right)\frac{1}{(1+\xi\,x^2)^2}- \frac{z}{2}f_0(\xi)\right],
\eea
and
\bea
\beta_2&=&  i\hat{\nu}\, m_D^2 \int \frac{d\Omega'}{4\pi} \left(\frac{z}{z-\bm{\hat{p}}\cdot\bm{v}'}-1\right) \int \frac{d\Omega}{4\pi} \frac{\bm{\hat{p}}\cdot\bm{v}+\xi \, t\, \cos\theta}{(1+\xi \cos^2\theta)^2} \frac{\hat{\omega}}{\hat{\omega}+i\hat{\nu}-\bm{\hat{p}}\cdot\bm{v}} \mathcal{W}(\hat{\omega},\hat{\nu})^{-1} \nn \\
&=&\frac{\left[z-z\mathcal{W}(\hat{\omega},\hat{\nu})-i\hat{\nu}\right]\hat{\omega}}{\mathcal{W}(\hat{\omega},\hat{\nu})} \left[\frac{z}{2}\mathcal{I}^{[0]}(\xi,\hat{\omega},t,\hat{\nu})+\frac{\xi \, t}{2}\mathcal{I}^{[1]}(\xi,\hat{\omega},t,\hat{\nu})-\frac{f_0(\xi)}{2}\right].
\eea
Combining the above two equations, we get,
\bea
 \beta(\xi,\hat{\omega},t,\hat{\nu})&=&- \frac{\hat{\omega}^2m_D^2}{2\mathcal{W}(\hat{\omega},\hat{\nu})}[f_0 (\xi)-z \mathcal{I}^{[0]}(\xi,\hat{\omega},t,\hat{\nu})-\xi t \mathcal{I}^{[1]}(\xi,\hat{\omega},t,\hat{\nu})].
\eea
Now, from Eq.~\eqref{struc_func}, we can obtain 
\be\label{alpha_def}
\alpha=\text{Tr}\Pi^{ij}-\frac{1}{\Tilde{n}^2}\Tilde{n}^i \Pi^{ij}\Tilde{n}^j-\beta,
\ee
with
\bea
\text{Tr} \Pi^{ij} &=& \Pi^{ii} \nn \\
&=&\frac{m_D^2}{2}\left[\xi f_1(\xi)+ \hat{\omega}\,\mathcal{I}^{[0]}(\xi,\hat{\omega},t,\hat{\nu})+\left(\xi \, t - \frac{i\hat{\nu}(\hat{\omega}-i\xi\,\hat{\nu})\left(1- \frac{z}{2}\log \frac{z+1}{z-1}\right)}{1-\frac{i\hat{\nu}}{2}\log\frac{z+1}{z-1}}\right) \mathcal{I}^{[1]}(\xi,\hat{\omega},t,\hat{\nu})\right. \nn \\
&&- \left.i\hat{\nu}\xi t\left(\frac{1}{t} +  \ \frac{1- \frac{z}{2}\log \frac{z+1}{z-1}}{1-\frac{i\hat{\nu}}{2}\log\frac{z+1}{z-1}}\right) \mathcal{I}^{[2]}(\xi,\hat{\omega},t,\hat{\nu})\right].
\eea
Now, in Eq.~\eqref{alpha_def}, before obtaining the second term, note that, $\Tilde{n}^i=A^{ij}n^j=n^i-t\, \hat{p}^i$, and $\Tilde{n}^2=1-t^2$.
\bea
\frac{1}{\Tilde{n}^2}\Tilde{n}^i \Pi^{ij}\Tilde{n}^j 
&=& \frac{m_D^2}{1-t^2}\bigg[-\frac{t^2\hat{\omega}z}{2}f_0(\xi)+ \frac{\xi}{2}f_1(\xi)+\frac{1}{2}t^2\,\hat{\omega}z^2 \,\mathcal{I}^{[0]}(\xi,\hat{\omega},t,\hat{\nu})+ \frac{1}{2}tz(i\hat{\nu}\xi-2\hat{\omega}+t^2\xi \hat{\omega})\mathcal{I}^{[1]}(\xi,\hat{\omega},t,\hat{\nu}) \nn \\
&+&\frac{1}{2}(-i(1+t^2)\hat{\nu}\xi+\hat{\omega}-2t^2\xi\hat{\omega})\mathcal{I}^{[2]}(\xi,\hat{\omega},t,\hat{\nu})+\frac{1}{2}\xi\, t \mathcal{I}^{[3]}(\xi,\hat{\omega},t,\hat{\nu})\bigg]+m_D^2\frac{i\hat{\nu}(t-1)}{1-t^2}\frac{2-z\log\frac{z+1}{z-1}}{2-i\hat{\nu}\log\frac{z+1}{z-1}} \nn \\
&\times& \bigg[ \frac{\hat{\omega}t}{2} f_0(\xi)- \frac{\hat{\omega}t\, z}{2}\mathcal{I}^{[0]}(\xi,\hat{\omega},t,\hat{\nu})+\frac{\hat{\omega}-\xi \,t^2 \, z-i\hat{\nu}\, \xi \, (1-t^2)}{2}\mathcal{I}^{[1]}(\xi,\hat{\omega},t,\hat{\nu})+\frac{\xi\, t}{2}\mathcal{I}^{[2]}(\xi,\hat{\omega},t,\hat{\nu}) \bigg].
\eea
So, Eq.~\eqref{alpha_def} becomes
\bea
 \alpha(\xi, \hat{\omega},t,\hat{\nu})&=& \frac{m_D^2}{2(1-t^2)}\left[\hat{\omega}zf_0(\xi)-\xi t^2 f_1(\xi)+\hat{\omega}(1-t^2-z^2)\mathcal{I}^{[0]}(\xi,\hat{\omega},t,\hat{\nu}) +t\left\{\xi (1-t^2)+2\hat{\omega}z-\xi z^2\right\}\right.\nn \\
        &\times&\left. \mathcal{I}^{[1]}(\xi,\hat{\omega},t,\hat{\nu})-(\hat{\omega}-2 \xi z t^2)\mathcal{I}^{[2]}(\xi,\hat{\omega},t,\hat{\nu})-\xi t \mathcal{I}^{[3]}(\xi,\hat{\omega},t,\hat{\nu})\right].
\eea
Now, the structure function $\gamma(\xi,\hat{\omega},t,\hat{\nu})$ can be obtained as
\bea
\gamma(\xi,\hat{\omega},t,\hat{\nu})&=& \frac{1}{\Tilde{n}^2}\Tilde{n}^i \Pi^{ij}\Tilde{n}^j- \alpha(\xi,\hat{\omega},t,\hat{\nu}) \nn \\
&=& \frac{m_D^2}{2(1-t^2)}\bigg[\xi (1+t^2)f_1(\xi)-\hat{\omega}z (1+t^2) f_0(\xi)-\hat{\omega}((1-t^2)-z^2(1+t^2))\mathcal{I}^{[0]}(\xi,\hat{\omega},t,\hat{\nu})\nn \\
        &- &t(4\hat{\omega}z -2 \xi z^2+\xi(1-t^2)(1+\hat{\omega}z))\mathcal{I}^{[1]}(\xi,\hat{\omega},t,\hat{\nu}) \nn \\
        &-& (4\xi z- 2\hat{\omega}-\xi (1-t^2)(\hat{\omega}+3z))\mathcal{I}^{[2]}(\xi,\hat{\omega},t,\hat{\nu})+2\xi t \mathcal{I}^{[3]}(\xi,\hat{\omega},t,\hat{\nu})\bigg]
\eea
From Eq.~\eqref{struc_func}, $\rho(\xi,\hat{\omega},t,\hat{\nu})$ can be written as,
\bea
\rho &=& m_D^2\frac{\hat{p}^i\,\Tilde{n}^j}{\Tilde{n}^2}\bigg[ \int \frac{d\Omega}{4\pi} v^i \, \frac{v^l+\xi\, (\bm{v}\cdot\bm{n})n^l}{(1+\xi \, (\bm{v}\cdot\bm{n})^2)^2} \frac{\delta^{lj}(\hat{\omega}-\bm{\hat{p}}\cdot\bm{v})+\hat{p}^l\,v^j}{\hat{\omega}+i\,\hat{\nu}-\bm{\hat{p}}\cdot\bm{v}} + i\hat{\nu}\int \frac{d\Omega'}{4\pi} \frac{v'^{i}}{\hat{\omega}+i\,\hat{\nu}-\bm{\hat{p}}\cdot\bm{v'}} \nn \\ 
&\times& \int \frac{d\Omega}{4\pi}\frac{v^l+\xi\, (\bm{v}\cdot\bm{n})n^l}{(1+\xi \, (\bm{v}\cdot\bm{n})^2)^2} \frac{\delta^{lj}(\hat{\omega}-\bm{\hat{p}}\cdot\bm{v})+\hat{p}^l\,v^j}{\hat{\omega}+i\,\hat{\nu}-\bm{\hat{p}}\cdot\bm{v}} \frac{1}{\mathcal{W}(\hat{\omega},\hat{\nu})} \bigg] \nn \\
&=& \rho_1+ \rho_2,
\eea
with
\bea
\rho_1&=& m_D^2\frac{1}{\Tilde{n}^2}\int \frac{d\Omega}{4\pi} (\bm{\hat{p}}\cdot\bm{v})\frac{v^l+\xi\, (\bm{v}\cdot\bm{n})n^l}{(1+\xi\, (\bm{v}\cdot\bm{n})^2)^2} \frac{\Tilde{n}^l(\hat{\omega}-\bm{\hat{p}}\cdot\bm{v})+\hat{p}^l(\bm{\Tilde{n}}\cdot\bm{v})}{\hat{\omega}+i\hat{\nu}-\bm{\hat{p}}\cdot\bm{v}} \nn \\
&=& \frac{m_D^2}{2(1-t^2)}[\xi\, t\, z\, \mathcal{I}^{[2]}(\xi,\hat{\omega},t,\hat{\nu})- \hat{\omega}\,t\, z^2 \, \mathcal{I}^{[0]}(\xi,\hat{\omega},t,\hat{\nu})+z\, (\hat{\omega}-\xi\, t^2\, z-i\hat{\nu}\, \xi\, (1-t^2)) \nn \\
&& \times \mathcal{I}^{[1]}(\xi,\hat{\omega},t,\hat{\nu})+\hat{\omega}\,t\, z \, f_0(\xi)]\label{rho1}
\eea
and
\bea
\rho_2 &=& \frac{i\hat{\nu} m_D^2}{1-t^2}\int \frac{d\Omega'}{4\pi} \frac{\bm{\hat{p}}\cdot\bm{v'}}{\hat{\omega}+i\hat{\nu}-\bm{\hat{p}}\cdot\bm{v'}} \int \frac{d\Omega}{4\pi} \frac{v^l+\xi\, (\bm{v}\cdot\bm{n})n^l}{(1+\xi\, (\bm{v}\cdot\bm{n})^2)^2}\frac{\Tilde{n}^l+(\hat{\omega}-\bm{\hat{p}}\cdot\bm{v})+\hat{p}^l\, (\bm{v}\cdot\bm{\Tilde{n}})}{(\hat{\omega}+i\hat{\nu}-\bm{\hat{p}}\cdot\bm{v}) \mathcal{W}(\hat{\omega},\hat{\nu})} \nn \\
&=&m_D^2 \frac{\hat{\omega}-z\, \mathcal{W}(\hat{\omega},\hat{\nu})}{2(1-t^2)\mathcal{W}(\hat{\omega},\hat{\nu})} [ \hat{\omega}\, t\, f_0(\xi)+ \xi\, t\, \mathcal{I}^{[2]}(\xi,\hat{\omega},t,\hat{\nu})- \hat{\omega}t\,z \mathcal{I}^{[0]}(\xi,\hat{\omega},t,\hat{\nu})+ (\hat{\omega}-i\hat{\nu}\xi(1-t^2)-\xi, t^2\,z) \nn \\
&&\times\ \mathcal{I}^{[1]}(\xi,\hat{\omega},t,\hat{\nu})]\label{rho2}.
\eea
Combining Eqs.~\eqref{rho1}~and~\eqref{rho2}, we get
\bea
\rho(\xi,\hat{\omega},t,\hat{\nu})&=&\frac{\hat{\omega}m_D^2}{2\mathcal{W}(\hat{\omega},\hat{\nu})(1-t^2)}\left[\hat{\omega} t f_0(\xi)-\hat{\omega}t z \mathcal{I}^{[0]}(\xi,\hat{\omega},t,\hat{\nu})\right.\nn\\
&&\left.+\left\{\hat{\omega}(1+\xi -\xi t^2)-\xi z\right\}\mathcal{I}^{[1]}(\xi,\hat{\omega},t,\hat{\nu}) 
        +\xi t \mathcal{I}^{[2]}(\xi,\hat{\omega},t,\hat{\nu})\right].
\eea
Finally, we can get the expression of $\delta(\xi,\hat{\omega},t,\hat{\nu})$ from Eq.~\eqref{struc_func} as
\bea
\delta &=& m_D^2\frac{\hat{p}^j\,\Tilde{n}^i}{\Tilde{n}^2}\bigg[ \int \frac{d\Omega}{4\pi} v^i \, \frac{v^l+\xi\, (\bm{v}\cdot\bm{n})n^l}{(1+\xi \, (\bm{v}\cdot\bm{n})^2)^2} \frac{\delta^{lj}(\hat{\omega}-\bm{\hat{p}}\cdot\bm{v})+\hat{p}^l\,v^j}{\hat{\omega}+i\,\hat{\nu}-\bm{\hat{p}}\cdot\bm{v}} + i\hat{\nu}\int \frac{d\Omega'}{4\pi} \frac{v'^{i}}{\hat{\omega}+i\,\hat{\nu}-\bm{\hat{p}}\cdot\bm{v'}} \nn \\ 
&\times& \int \frac{d\Omega}{4\pi}\frac{v^l+\xi\, (\bm{v}\cdot\bm{n})n^l}{(1+\xi \, (\bm{v}\cdot\bm{n})^2)^2} \frac{\delta^{lj}(\hat{\omega}-\bm{\hat{p}}\cdot\bm{v})+\hat{p}^l\,v^j}{\hat{\omega}+i\,\hat{\nu}-\bm{\hat{p}}\cdot\bm{v}} \frac{1}{\mathcal{W}(\hat{\omega},\hat{\nu})} \bigg] \nn \\
&=& \frac{\hat{\omega}m_D^2}{2(1-t^2)}\Big[tzf_0(\xi)-tz^2 \mathcal{I}^{[0]}(\xi,\hat{\omega},t,\hat{\nu})+z(1-\xi t^2)\mathcal{I}^{[1]}(\xi,\hat{\omega},t,\hat{\nu})+\xi t \mathcal{I}^{[2]}(\xi,\hat{\omega},t,\hat{\nu})\Big] .
\eea

\section{Tensor relation}
\label{app:tensor_relation}

In TAG, the relevant propagator components are the
spatial components $\Delta^{ij}$. The inverse dielectric function and the
fluctuation function are therefore obtained from the longitudinal
projection of the spatial propagator,
\be
\Delta_L(P)\equiv \hat p^i \Delta^{ij}(P)\hat p^j .
\ee
Accordingly, in the static limit,
\be
\epsilon^{-1}(p)=-\lim_{\omega\to0}\omega^2
\hat p^i\Delta^{ij}(\omega,\bm p)\hat p^j .
\ee

The spatial components of the basis tensors are
\bea
A^{ij}&=&\delta^{ij}-\hat p^i\hat p^j, \qquad
B^{ij}=\hat p^i\hat p^j, \qquad
C^{ij}= \frac{\Tilde n^i\Tilde n^j}{\Tilde n^2}, \nn\\
D^{ij}&=&\Tilde n^i\hat p^j, \qquad
E^{ij}=\hat p^i\Tilde n^j ,
\eea
where
\be
\Tilde n^i \equiv A^{ij}n^j
= n^i-(\bm n\cdot\hat{\bm p})\hat p^i ,
\qquad
\Tilde n^2=1-t^2=\frac{p_\perp^2}{p^2}.
\ee
The vector $\Tilde n^i$ is orthogonal to $\hat p^i$.

The useful contraction identities are
\begin{subequations}
\bea
A^{ij}A^{ij}&=&2, \qquad
B^{ij}B^{ij}=1, \qquad
C^{ij}C^{ij}=1, \nn\\
D^{ij}D^{ij}&=&\Tilde n^2, \qquad
E^{ij}E^{ij}=\Tilde n^2,
\eea
\bea
\hat p^iA^{ij}&=&0, \qquad
\hat p^iB^{ij}=\hat p^j, \qquad
\hat p^iC^{ij}=0, \nn\\
\hat p^iD^{ij}&=&0, \qquad
\hat p^iE^{ij}= \Tilde n^j,
\eea
\bea
\Tilde n^iA^{ij}&=&\Tilde n^j, \qquad
\Tilde n^iB^{ij}=0, \qquad
\Tilde n^iC^{ij}=\Tilde n^j, \nn\\
\Tilde n^iD^{ij}&=&\Tilde n^2\hat p^j, \qquad
\Tilde n^iE^{ij}=0 .
\eea
\end{subequations}
For a general tensor decomposed as
\be
M^{ij}=M_A A^{ij}+M_B B^{ij}+M_C C^{ij}
+M_D D^{ij}+M_E E^{ij},
\ee
the longitudinal TAG projection gives
\be
\hat p^i M^{ij}\hat p^j=M_B .
\ee
Therefore, for the resummed gluon propagator
\be
\Delta^{ij}
=\alpha' A^{ij}+\beta' B^{ij}+\gamma' C^{ij}
+\delta' D^{ij}+\rho' E^{ij},
\ee
one obtains
\be
\hat p^i\Delta^{ij}\hat p^j=\beta' .
\ee
This relation is used in the main text to write
\be
\epsilon^{-1}(p)=-\lim_{\omega\to0}\omega^2\beta' .
\ee
For the Feynman propagator contribution, the required TAG projection is
\be
\left[\Delta_R\Pi_F\Delta_A\right]_{\rm long}
\equiv
\hat p^i\Delta_R^{ik}\Pi_F^{kl}\Delta_A^{lj}\hat p^j .
\ee
Using the tensor decomposition, the leading terms in the static limit are
\bea
\left[\Delta_R\Pi_F\Delta_A\right]_{\rm long}
&=&
\beta'_R\beta_F\beta'_A
+\rho'_R(\alpha_F+\gamma_F)\delta'_A\,\tilde n^2 .
\eea
Since $\tilde n^2=p_\perp^2/p^2$, one obtains
\be
\lim_{\omega\to0}\omega^2
\left[\Delta_R\Pi_F\Delta_A\right]_{\rm long}
=
\omega^2\beta'_R\beta_F\beta'_A
+
\omega^2\rho'_R(\alpha_F+\gamma_F)\delta'_A
\frac{p_\perp^2}{p^2}.
\label{RFA_appen_B}
\ee
This is the expression used in Sec.~\ref{sec:eps} for the calculation of the fluctuation function in the temporal axial gauge.

\section{Calculations of structure functions for the symmetric propagator}
\label{struc_fun_F}

The retarded and advanced self-energies derived in Sec.~\ref{sec:prop} are insufficient to fix the imaginary part of the in-medium potential, which requires the Feynman (symmetric) gluon propagator and, in turn,
the symmetric self-energy $\Pi_F^{ij}$. In an isotropic plasma, whether in or out of equilibrium and with or without collisions, the symmetric self-energy is fixed by the FDT once
the retarded self-energy is known. The FDT fails, however, as soon as the parton distribution becomes anisotropic, since no single
temperaturelike scale then relates the symmetric and retarded correlators. In that regime, $\Pi_F^{ij}$ must be obtained directly from the microscopic current-current correlator. We do so here within the conserving BGK kinetic framework, using the same notation as the main text: $P=(\omega,\bm{p})$ denotes the external four-momentum, $\bm{k}$ the parton phase-space momentum, and $\bm{v}=\bm{k}/k$ the corresponding velocity. The scale $\lambda$ specifies the typical momentum of the plasma particles and should be understood as the temperature only in the thermal-equilibrium limit; it sets the overall strength of the noise correlator and of $\Pi_F^{ij}$ derived below.

\subsection{Microscopic derivation of the Feynman self-energy from current correlators}
We extend the kinetic theory of Sec.~\ref{sec:prop} by promoting the
phase-space fluctuation $\delta f(\bm{x},\bm{v},t)$ to a stochastic
variable.  Introducing a local noise source $\zeta(\bm{x},\bm{v},t)$
that represents the microscopic fluctuations not captured by the
averaged BGK kernel, the linearized Boltzmann equation
(\ref{boltzmann_eq}) acquires the form
\bea\label{stoch_boltz}
\bigl(\partial_t + \bm{v}\!\cdot\!\bm{\nabla} + \nu\bigr)\,
\delta f(\bm{x},\bm{v},t)
= \nu\,\frac{\delta n(\bm{x},t)}{\langle 1\rangle_\xi}\,
f(\bm{v}) + \zeta(\bm{x},\bm{v},t),
\eea
where $f(\bm{v})$ is the anisotropic distribution (\ref{f_dist}) expressed
in terms of the velocity direction, and the conserving subtraction
$\propto \delta n/\langle 1\rangle_\xi$ enforces local
particle-number conservation.  The introduction of the stochastic noise source $\zeta$ allows us to 
systematically deduce the symmetric self-energy $\Pi_F^{ij}$ and the 
resulting propagator from microscopic fluctuations without prematurely assuming a FDT relation. Crucially, in the limit where the anisotropy vanishes ($\xi \to 0$), the noise correlations dictate that the stochastic framework naturally reduces to the expected FDT relation for a collisional isotropic plasma. For $\xi \neq 0$, it successfully and dynamically captures the nontrivial, 
anisotropy-induced modifications to the symmetric propagator that are inherently missing from a deterministic transport equation.

The anisotropic angular average of a velocity-dependent quantity $X$ is defined, without normalization, as
\be\label{avg_def}
\langle X\rangle_\xi \;\equiv\;
\int\!\frac{d\Omega_v}{4\pi}\,
\frac{X(\bm{v})}{\bigl[1+\xi(\bm{v}\!\cdot\!\bm{n})^2\bigr]^{3/2}},
\ee
so that $\langle 1\rangle_\xi$ is in general different from unity and
is kept explicit in every expression below.  The deformation weight
$(1+\xi(\bm{v}\!\cdot\!\bm{n})^2)^{-3/2}$ follows from the Jacobian of
the spheroidal change of variables
$k'\!=\!k\sqrt{1+\xi(\bm{v}\!\cdot\!\bm{n})^2}$ that maps the
anisotropic distribution onto an isotropic one; the radial momentum integral has already been absorbed into $m_D^2$ through
Eq.~(\ref{debyemass}).

\paragraph{Noise correlator:}
Local conservation requires $\int\! d^3k/(2\pi)^3\,\zeta=0$, which
forces the noise to be anticorrelated with its own angular average.
At the effective scale $\lambda$, the conserving two-point function reads
\bea\label{noise_corr}
\big\langle \zeta(P,\bm{v})\,\zeta^*(P',\bm{v}')\big\rangle_{\rm sto}
&=& 2\nu\,\lambda\,m_D^2\,(2\pi)^4\,\delta^{(4)}(P\!-\!P')\,
\frac{(2\pi)^3}{k^2}
\left(-\frac{\partial f_{\rm iso}}{\partial k}\right)  \nn\\[2pt]
&&\times
\left[
\delta(\bm{v}-\bm{v}')
- \frac{\bigl[1+\xi(\bm{v}\!\cdot\!\bm{n})^2\bigr]^{-3/2}}
{\langle 1\rangle_\xi}
\right],
\eea
which is constructed so that
$\int\! d\Omega_v/(4\pi)\,\langle\zeta(P,\bm{v})\zeta^*(P',\bm{v}')
\rangle_{\rm sto}=0$.
In the isotropic limit $\langle 1\rangle_\xi\to1$, and with
$\lambda\to T$ in thermal equilibrium the amplitude $2\nu\lambda m_D^2$
reduces to the standard Langevin strength $2\nu T m_D^2$ of the
fluctuation-dissipation relation.

\paragraph{Solution for $\delta f$.}
Fourier transforming Eq.~(\ref{stoch_boltz}) with $P=(\omega,\bm{p})$
gives
\be\label{df_intermediate}
\delta f(P,\bm{v}) =
\frac{i\,\nu\,(-\partial f_0/\partial k)}
{\omega-\bm{v}\!\cdot\!\bm{p}+i\nu}\,
\langle\delta f(P,\bm{v})\rangle_\xi
+ \frac{i\,\zeta(P,\bm{v})}{\omega-\bm{v}\!\cdot\!\bm{p}+i\nu}.
\ee
Applying $\langle\cdots\rangle_\xi$ to both sides and using the conserving property of the noise, one obtains
\be\label{bracket_avg}
\langle\delta f\rangle_\xi
\left[1 - \frac{i\nu}{\langle 1\rangle_\xi}
\left\langle\frac{1}{\omega-\bm{v}\!\cdot\!\bm{p}+i\nu}
\right\rangle_\xi\right]
= i\left\langle\frac{\zeta}{\omega-\bm{v}\!\cdot\!\bm{p}+i\nu}
\right\rangle_\xi,
\ee
where the bracket on the left defines the conserving denominator
\be\label{D_def}
D(\omega,\bm{p}) \;\equiv\;
1 - \frac{i\nu}{\langle 1\rangle_\xi}
\left\langle\frac{1}{\omega-\bm{v}\!\cdot\!\bm{p}+i\nu}
\right\rangle_\xi,
\ee
which reduces, in the isotropic limit, to the function
$\mathcal{W}(\hat\omega,\hat\nu)$ of Eq.~(\ref{self energy}) via
$D|_{\xi\to0}=\mathcal{W}$.  Substituting Eq.~(\ref{bracket_avg}) back
into Eq.~(\ref{df_intermediate}) gives the closed-form fluctuation
\be\label{final_df}
\delta f(P,\bm{v}) =
\frac{i\,\zeta(P,\bm{v})}{\omega-\bm{v}\!\cdot\!\bm{p}+i\nu}
- \frac{\nu\,(-\partial f_0/\partial k)}
{\omega-\bm{v}\!\cdot\!\bm{p}+i\nu}\,
\frac{1}{D(\omega,\bm{p})}\,
\left\langle\frac{\zeta(P,\bm{v})}{\omega-\bm{v}\!\cdot\!\bm{p}+i\nu}
\right\rangle_\xi .
\ee
Equation~(\ref{final_df}) decomposes the total fluctuation into a
direct noise-driven piece (first term) and a backreaction piece
(second term) required by local conservation; both are linear and homogeneous in $\zeta$.

\paragraph{Color current and symmetric self-energy.}
The fluctuating color current is the stochastic piece of the induced
current $J^{\mu}_{\rm ind,a}(P)$ of Eq.~\eqref{J_ind}; it is obtained by inserting the noise-driven fluctuation (\ref{final_df}) into the
same current functional,
\be\label{delta_j_def}
\delta j^i(P) \;\equiv\; g
\int_{\bm{k}} V^{\,i}\,
\bigl\{2N_c\,\delta f^g_a(P,\bm{v})
+ N_f\bigl[\delta f^q_a(P,\bm{v})-\delta f^{\bar q}_a(P,\bm{v})
\bigr]\bigr\},
\ee
where $\int_{\bm{k}}\equiv\int d^3k/(2\pi)^3$ and $V^\mu=(1,\bm{v})$.
The functional form of $\delta j^i$ is identical to that of
$J^{\mu}_{\rm ind,a}$; the difference is solely in the input: in the
retarded response, $\delta f$ is sourced by the gauge field $\delta A$,
whereas here it is sourced by the noise $\zeta$ through
Eq.~(\ref{final_df}).  The retarded self-energy is the linear response
$\Pi_R^{\mu\nu}=\delta J^\mu_{\rm ind}/\delta A_\nu$, whereas the symmetric self-energy is the statistical two-point correlator of the
same current, evaluated on the closed Schwinger-Keldysh contour. So, one gets
\be\label{Pi_F_def}
\Pi_F^{ij}(\omega,\bm{p}) \;\equiv\;
-\,2 i\;
\big\langle \delta j^i(\omega,\bm{p})\,
\delta j^{j}(-\omega,-\bm{p})\big\rangle_{\rm sto},
\ee
which is the quantity entering the fluctuation function
$\Phi(p,\xi,\nu)$ of Sec.~\ref{sec:eps}.

\paragraph{Direct and vertex contributions.}
It is useful to separate Eq.~\eqref{final_df} into the two pieces
\bea
\delta f(P,\bm v)
&=&
\delta f_{\rm dir}(P,\bm v)+\delta f_{\rm ver}(P,\bm v),
\eea
where
\bea
\delta f_{\rm dir}(P,\bm v)
&=&
\frac{i\,\zeta(P,\bm v)}
{\omega-\bm v\cdot\bm p+i\nu},
\\
\delta f_{\rm ver}(P,\bm v)
&=&
-\frac{\nu\,(-\partial f_0/\partial k)}
{\omega-\bm v\cdot\bm p+i\nu}
\frac{1}{D(\omega,\bm p)}
\left\langle
\frac{\zeta(P,\bm v)}
{\omega-\bm v\cdot\bm p+i\nu}
\right\rangle_\xi .
\eea
Correspondingly, the fluctuating current can be written as
\be
\delta j^i(P)=\delta j_{\rm dir}^i(P)+\delta j_{\rm ver}^i(P),
\ee
with
\bea
\delta j_{\rm dir}^i(P)
&=&
g\int_{\bm k}v^i\,\delta f_{\rm dir}(P,\bm v),
\\
\delta j_{\rm ver}^i(P)
&=&
g\int_{\bm k}v^i\,\delta f_{\rm ver}(P,\bm v).
\eea
Suppressing the common color and degeneracy factors, which are already absorbed into the definition of $m_D^2$, the symmetric self-energy is defined by
\be
\Pi_F^{ij}(\omega,\bm p)
=
-2 i\,
\left\langle
\delta j^i(\omega,\bm p)\,
\delta j^j(-\omega,-\bm p)
\right\rangle_{\rm sto}.
\label{Pi_F_def_again}
\ee
Therefore,
\bea
\Pi_F^{ij}
&=&
\Pi_{F,\rm direct}^{ij}
+\Pi_{F,\rm vertex}^{ij},
\eea
where the direct contribution is generated by
$\langle\delta j_{\rm dir}^i\delta j_{\rm dir}^j\rangle_{\rm sto}$,
while the vertex contribution collects the terms induced by the conserving BGK subtraction,
\bea
\Pi_{F,\rm vertex}^{ij}
\equiv
-2 i\,\left[
\left\langle\delta j_{\rm dir}^i\delta j_{\rm ver}^j\right\rangle_{\rm sto}
+ \left\langle\delta j_{\rm ver}^i\delta j_{\rm dir}^j\right\rangle_{\rm sto} +
\left\langle\delta j_{\rm ver}^i\delta j_{\rm ver}^j\right\rangle_{\rm sto} \right].
\eea
Using the noise correlator in Eq.~\eqref{noise_corr}, the direct part becomes
\bea
\Pi_{F,\rm direct}^{ij}
&=&
-2 i\,
\left\langle\delta j_{\rm dir}^i(P)\delta j_{\rm dir}^j(-P)\right\rangle_{\rm sto}
\nn\\
&=&-2 i\,\left[
2\nu\,\lambda\,m_D^2
\left\langle
\frac{v^i v^j}
{(\omega-\bm v\cdot\bm p)^2+\nu^2}
\right\rangle_\xi
\right].
\label{Pi_F_direct}
\eea
Here we have used
\be
\frac{i}{\omega-\bm v\cdot\bm p+i\nu} \,
\frac{i}{-\omega+\bm v\cdot\bm p+i\nu}
=
-\frac{1}{(\omega-\bm v\cdot\bm p)^2+\nu^2},
\ee
and the sign is included in the overall Schwinger-Keldysh factor
$-2\pi i$ in Eq.~\eqref{Pi_F_def_again}.

The conserving-subtraction terms give
\bea
\Pi_{F,\rm vertex}^{ij}
&=&
-2 i\,
\left[
\left\langle\delta j_{\rm dir}^i\delta j_{\rm ver}^j\right\rangle_{\rm sto}
+
\left\langle\delta j_{\rm ver}^i\delta j_{\rm dir}^j\right\rangle_{\rm sto}
+
\left\langle\delta j_{\rm ver}^i\delta j_{\rm ver}^j\right\rangle_{\rm sto}
\right]
\nn\\
&=&
-2 i\,
\left[
-\frac{2\nu\,\lambda\,m_D^2}{\langle 1\rangle_\xi}\,
\mathrm{Re}\left\{
\frac{
\left\langle
\dfrac{v^i}{\omega-\bm v\cdot\bm p+i\nu}
\right\rangle_\xi
\left\langle
\dfrac{v^j}{\omega-\bm v\cdot\bm p+i\nu}
\right\rangle_\xi
}
{D(\omega,\bm p)}
\right\}
\right].
\label{Pi_F_vertex}
\eea
The factor $1/\langle 1\rangle_\xi$ originates from the conserving normalization in the BGK kernel. It is required because the angular
average in Eq.~\eqref{avg_def} is not normalized.

Thus, the full symmetric self-energy is
\bea
\Pi_F^{ij}(\omega,\bm p)
&=&
-4 i\,\nu\,\lambda\,m_D^2
\left[
\left\langle
\frac{v^i v^j}
{(\omega-\bm v\cdot\bm p)^2+\nu^2}
\right\rangle_\xi
\right.
\nn\\
&&
\left.
-
\frac{1}{\langle 1\rangle_\xi}
\mathrm{Re}\left\{
\frac{
\left\langle
\dfrac{v^i}{\omega-\bm v\cdot\bm p+i\nu}
\right\rangle_\xi
\left\langle
\dfrac{v^j}{\omega-\bm v\cdot\bm p+i\nu}
\right\rangle_\xi
}
{D(\omega,\bm p)}
\right\}
\right].
\label{full_pif}
\eea
The first term is the uncorrelated direct fluctuation, while the second term is the vertex correction required by local particle-number conservation in the BGK kernel.

\subsection{Form factors of the Feynman self-energy in the static limit}

Projecting Eq.~\eqref{full_pif} onto the tensor basis of Sec.~\ref{sec:prop} --- the longitudinal projector
$\hat{\bm p}=\bm{p}/p$ and the orthogonalized anisotropy vector $\tilde{\bm n}=\bm{n}-(\bm{n}\!\cdot\!\hat{\bm p})\hat{\bm p}$ with
$\tilde n^2=1-t^2=p_\perp^2/p^2$ --- fixes the Feynman structure functions through the contractions of Eq.~\eqref{struc_func}. In the
static limit $\omega\to 0$, the leading nonvanishing pieces are
\bea\label{aF_static}
\lim_{\omega\to0}\alpha_F
&=& -\frac{4i\hat\nu\,\lambda\,m_D^2}{p}\,
\left\langle
\frac{1-\dfrac{(\bm{v}\!\cdot\!\tilde{\bm n})^2}{\tilde n^2}}
{(\bm{v}\!\cdot\!\hat{\bm p})^2+\hat\nu^2}
\right\rangle_\xi ,\\[6pt]
\label{bF_static}
\lim_{\omega\to0}\beta_F
&=& -\frac{4i\,\lambda\,m_D^2\,\omega^2}{p^3}\,
\frac{\langle 1\rangle_\xi
\bigl\langle\dfrac{\hat\nu}
{(\bm{v}\!\cdot\!\hat{\bm p})^2+\hat\nu^2}\bigr\rangle_\xi}
{\bigl\langle\dfrac{(\bm{v}\!\cdot\!\hat{\bm p})^2}
{(\bm{v}\!\cdot\!\hat{\bm p})^2+\hat\nu^2}\bigr\rangle_\xi},
\\[6pt]
\label{gF_static}
\lim_{\omega\to0}\gamma_F
&=& -\frac{4i\hat\nu\,\lambda\,m_D^2}{p}\,
\left\langle
\frac{\dfrac{2(\bm{v}\!\cdot\!\tilde{\bm n})^2}{\tilde n^2}-1}
{(\bm{v}\!\cdot\!\hat{\bm p})^2+\hat\nu^2}
\right\rangle_\xi .
\eea
The off-diagonal structure functions satisfy the symmetry
$\delta_F=\rho_F$ (a consequence of the symmetric $ij$ structure of
$\Pi_F$, in contrast to the asymmetric retarded self-energy), and are given in the static limit by
\bea\label{dF_static}
\lim_{\omega\to0}\delta_F
&=& \lim_{\omega\to0}\rho_F\nn\\
&=& -\frac{4i\hat\nu\,\lambda\,m_D^2\,\omega^2}{p^3\,\tilde n^2}\,
\frac{\left\langle 1\right\rangle_\xi}
{\left\langle\dfrac{(\bm{v}\!\cdot\!\hat{\bm p})^2}
{\hat\nu^2+(\bm{v}\!\cdot\!\hat{\bm p})^2}\right\rangle_\xi}
\left[
\frac{
\left\langle\dfrac{(\bm{v}\!\cdot\!\tilde{\bm n})(\bm{v}\!\cdot\!\hat{\bm p})}
{(\bm{v}\!\cdot\!\hat{\bm p})^2+\hat\nu^2}\right\rangle_\xi
\left\langle\dfrac{\hat\nu^2-(\bm{v}\!\cdot\!\hat{\bm p})^2}
{(\hat\nu^2+(\bm{v}\!\cdot\!\hat{\bm p})^2)^2}\right\rangle_\xi}
{\left\langle\dfrac{(\bm{v}\!\cdot\!\hat{\bm p})^2}
{\hat\nu^2+(\bm{v}\!\cdot\!\hat{\bm p})^2}\right\rangle_\xi}
+2\bigl\langle\dfrac{(\bm{v}\!\cdot\!\tilde{\bm n})(\bm{v}\!\cdot\!\hat{\bm p})}
{\bigl[(\bm{v}\!\cdot\!\hat{\bm p})^2+\hat\nu^2\bigr]^2}\bigr\rangle_\xi
\right].
\eea
These results demonstrate that, although the retarded self-energy is asymmetric in $i\leftrightarrow j$ because the BGK kernel breaks time-reversal symmetry, the Feynman self-energy retains the symmetric structure $\Pi_F^{ij}=\Pi_F^{ji}$ and hence $\delta_F=\rho_F$.

\subsection{Collisionless limit}

For $\nu\to0$, the Lorentzian kernel reduces, via
$\displaystyle\lim_{\nu\to0}\frac{\nu}{x^2+\nu^2}=\pi\,\delta(x)$,
to a Dirac delta that selects the resonant velocities
$\bm{v}\!\cdot\!\hat{\bm p}=\hat\omega$.  The angular integrals then evaluate in terms of the complete elliptic integrals
$K(\kappa)$ and $E(\kappa)$, with the effective anisotropy
$\varsigma\equiv\xi(1-t^2)=\xi\,\tilde n^2$ defined as in
Sec.~\ref{sec:eps}. The principal form factors become
\bea\label{aF_cl}
\lim_{\nu\to0}\alpha_F
&=& -\frac{4i\,\lambda\,m_D^2}{p\,\varsigma}\,
\bigl[E(-\varsigma)-K(-\varsigma)\bigr],\\[4pt]
\label{bF_cl}
\lim_{\nu\to0}\beta_F
&=& -\frac{4i\,\lambda\,m_D^2\,\omega^2}{p^3}\,
\frac{E(-\varsigma)}{1+\varsigma},\\[4pt]
\label{gF_cl}
\lim_{\nu\to0}\gamma_F
&=& -\frac{4i\,\lambda\,m_D^2}{p\,\varsigma}\,
\left[2K(-\varsigma)-\frac{2+\varsigma}{1+\varsigma}\,
E(-\varsigma)\right],
\eea
and the off-diagonal functions
\be\label{dF_cl}
\lim_{\nu\to0}\delta_F
= \lim_{\nu\to0}\rho_F
= -\frac{4i\,\lambda\,m_D^2\,\omega^2\,t}
{p^3\,(1-t^2)\,(1+\varsigma)}
\left[
\frac{1-\varsigma}{1+\varsigma}\,E(-\varsigma)-K(-\varsigma)
\right].
\ee
The fluctuations are nonzero only in the subluminal regime $|\omega|<p$, modified by the alignment of $\bm{p}$ relative to $\bm{n}$ through $\varsigma$, and reproduce the standard collisionless anisotropic HTL spectral functions.

\subsection{Isotropic limit}
As $\xi\to0$ the deformation weight collapses to unity, $\langle 1\rangle_\xi\to1$, and the angular averages reduce to plain spherical integrals. With $\langle(\bm{v}\!\cdot\!\hat{\bm p})^n\rangle_0 =\tfrac12\int_{-1}^{1}dx\,x^n$ and
$\langle x^2/(x^2+\hat\nu^2)\rangle_0
=1-\hat\nu\,\arccot\hat\nu$, the structure functions become
\bea\label{aF_iso}
\lim_{\xi\to0}\alpha_F
&=& -\frac{2i\,\lambda\,m_D^2}{p}\,
\bigl[(1-\hat\nu^2)\arccot\hat\nu+\hat\nu\bigr],\\[4pt]
\label{bF_iso}
\lim_{\xi\to0}\beta_F
&=& -\frac{4i\,\lambda\,m_D^2\,\omega^2}{p^3}\,
\frac{\arccot\hat\nu}{1-\hat\nu\,\arccot\hat\nu},\\[4pt]
\label{gF_iso}
\lim_{\xi\to0}\gamma_F
&=& -\frac{2i\,\lambda\,m_D^2}{p}\,
\bigl[3\hat\nu^2\arccot\hat\nu-3\hat\nu+\arccot\hat\nu\bigr].
\eea
The off-diagonal functions vanish identically, $\delta_F=\rho_F=0$,
because the anisotropy vector $\bm{n}$ is no longer defined and the
longitudinal/transverse projections decouple. In this limit the FDT is recovered, $\beta_F=(2\lambda/\omega)\,\mathrm{Im}\,\beta_R$, and with $\lambda\to T$ in thermal equilibrium, the result coincides with the standard isotropic collisional HTL expression, providing a nontrivial check on the conserving derivation above.


\begin{thebibliography}{47}%
\makeatletter
\providecommand \@ifxundefined [1]{%
 \@ifx{#1\undefined}
}%
\providecommand \@ifnum [1]{%
 \ifnum #1\expandafter \@firstoftwo
 \else \expandafter \@secondoftwo
 \fi
}%
\providecommand \@ifx [1]{%
 \ifx #1\expandafter \@firstoftwo
 \else \expandafter \@secondoftwo
 \fi
}%
\providecommand \natexlab [1]{#1}%
\providecommand \enquote  [1]{``#1''}%
\providecommand \bibnamefont  [1]{#1}%
\providecommand \bibfnamefont [1]{#1}%
\providecommand \citenamefont [1]{#1}%
\providecommand \href@noop [0]{\@secondoftwo}%
\providecommand \href [0]{\begingroup \@sanitize@url \@href}%
\providecommand \@href[1]{\@@startlink{#1}\@@href}%
\providecommand \@@href[1]{\endgroup#1\@@endlink}%
\providecommand \@sanitize@url [0]{\catcode `\\12\catcode `\$12\catcode `\&12\catcode `\#12\catcode `\^12\catcode `\_12\catcode `\%12\relax}%
\providecommand \@@startlink[1]{}%
\providecommand \@@endlink[0]{}%
\providecommand \url  [0]{\begingroup\@sanitize@url \@url }%
\providecommand \@url [1]{\endgroup\@href {#1}{\urlprefix }}%
\providecommand \urlprefix  [0]{URL }%
\providecommand \Eprint [0]{\href }%
\providecommand \doibase [0]{https://doi.org/}%
\providecommand \selectlanguage [0]{\@gobble}%
\providecommand \bibinfo  [0]{\@secondoftwo}%
\providecommand \bibfield  [0]{\@secondoftwo}%
\providecommand \translation [1]{[#1]}%
\providecommand \BibitemOpen [0]{}%
\providecommand \bibitemStop [0]{}%
\providecommand \bibitemNoStop [0]{.\EOS\space}%
\providecommand \EOS [0]{\spacefactor3000\relax}%
\providecommand \BibitemShut  [1]{\csname bibitem#1\endcsname}%
\let\auto@bib@innerbib\@empty
\bibitem [{\citenamefont {Weldon}(1982)}]{Weldon:1982aq}%
  \BibitemOpen
  \bibfield  {author} {\bibinfo {author} {\bibfnamefont {H.~A.}\ \bibnamefont {Weldon}},\ }\href {https://doi.org/10.1103/PhysRevD.26.1394} {\bibfield  {journal} {\bibinfo  {journal} {Phys. Rev. D}\ }\textbf {\bibinfo {volume} {26}},\ \bibinfo {pages} {1394} (\bibinfo {year} {1982})}\BibitemShut {NoStop}%
\bibitem [{\citenamefont {Braaten}\ and\ \citenamefont {Pisarski}(1990)}]{Braaten:1989mz}%
  \BibitemOpen
  \bibfield  {author} {\bibinfo {author} {\bibfnamefont {E.}~\bibnamefont {Braaten}}\ and\ \bibinfo {author} {\bibfnamefont {R.~D.}\ \bibnamefont {Pisarski}},\ }\href {https://doi.org/10.1016/0550-3213(90)90508-B} {\bibfield  {journal} {\bibinfo  {journal} {Nucl. Phys. B}\ }\textbf {\bibinfo {volume} {337}},\ \bibinfo {pages} {569} (\bibinfo {year} {1990})}\BibitemShut {NoStop}%
\bibitem [{\citenamefont {Frenkel}\ and\ \citenamefont {Taylor}(1990)}]{Frenkel:1989br}%
  \BibitemOpen
  \bibfield  {author} {\bibinfo {author} {\bibfnamefont {J.}~\bibnamefont {Frenkel}}\ and\ \bibinfo {author} {\bibfnamefont {J.~C.}\ \bibnamefont {Taylor}},\ }\href {https://doi.org/10.1016/0550-3213(90)90661-V} {\bibfield  {journal} {\bibinfo  {journal} {Nucl. Phys. B}\ }\textbf {\bibinfo {volume} {334}},\ \bibinfo {pages} {199} (\bibinfo {year} {1990})}\BibitemShut {NoStop}%
\bibitem [{\citenamefont {Braaten}\ and\ \citenamefont {Pisarski}(1992)}]{Braaten:1991gm}%
  \BibitemOpen
  \bibfield  {author} {\bibinfo {author} {\bibfnamefont {E.}~\bibnamefont {Braaten}}\ and\ \bibinfo {author} {\bibfnamefont {R.~D.}\ \bibnamefont {Pisarski}},\ }\href {https://doi.org/10.1103/PhysRevD.45.R1827} {\bibfield  {journal} {\bibinfo  {journal} {Phys. Rev. D}\ }\textbf {\bibinfo {volume} {45}},\ \bibinfo {pages} {R1827} (\bibinfo {year} {1992})}\BibitemShut {NoStop}%
\bibitem [{\citenamefont {Haque}\ \emph {et~al.}(2014)\citenamefont {Haque}, \citenamefont {Bandyopadhyay}, \citenamefont {Andersen}, \citenamefont {Mustafa}, \citenamefont {Strickland},\ and\ \citenamefont {Su}}]{Haque:2014rua}%
  \BibitemOpen
  \bibfield  {author} {\bibinfo {author} {\bibfnamefont {N.}~\bibnamefont {Haque}}, \bibinfo {author} {\bibfnamefont {A.}~\bibnamefont {Bandyopadhyay}}, \bibinfo {author} {\bibfnamefont {J.~O.}\ \bibnamefont {Andersen}}, \bibinfo {author} {\bibfnamefont {M.~G.}\ \bibnamefont {Mustafa}}, \bibinfo {author} {\bibfnamefont {M.}~\bibnamefont {Strickland}},\ and\ \bibinfo {author} {\bibfnamefont {N.}~\bibnamefont {Su}},\ }\href {https://doi.org/10.1007/JHEP05(2014)027} {\bibfield  {journal} {\bibinfo  {journal} {JHEP}\ }\textbf {\bibinfo {volume} {05}},\ \bibinfo {pages} {027}},\ \Eprint {https://arxiv.org/abs/1402.6907} {arXiv:1402.6907 [hep-ph]} \BibitemShut {NoStop}%
\bibitem [{\citenamefont {Haque}\ and\ \citenamefont {Mustafa}(2025)}]{Haque:2024gva}%
  \BibitemOpen
  \bibfield  {author} {\bibinfo {author} {\bibfnamefont {N.}~\bibnamefont {Haque}}\ and\ \bibinfo {author} {\bibfnamefont {M.~G.}\ \bibnamefont {Mustafa}},\ }\href {https://doi.org/10.1016/j.ppnp.2024.104136} {\bibfield  {journal} {\bibinfo  {journal} {Prog. Part. Nucl. Phys.}\ }\textbf {\bibinfo {volume} {140}},\ \bibinfo {pages} {104136} (\bibinfo {year} {2025})},\ \Eprint {https://arxiv.org/abs/2404.08734} {arXiv:2404.08734 [hep-ph]} \BibitemShut {NoStop}%
\bibitem [{\citenamefont {Matsui}\ and\ \citenamefont {Satz}(1986)}]{Matsui:1986dk}%
  \BibitemOpen
  \bibfield  {author} {\bibinfo {author} {\bibfnamefont {T.}~\bibnamefont {Matsui}}\ and\ \bibinfo {author} {\bibfnamefont {H.}~\bibnamefont {Satz}},\ }\href {https://doi.org/10.1016/0370-2693(86)91404-8} {\bibfield  {journal} {\bibinfo  {journal} {Phys. Lett. B}\ }\textbf {\bibinfo {volume} {178}},\ \bibinfo {pages} {416} (\bibinfo {year} {1986})}\BibitemShut {NoStop}%
\bibitem [{\citenamefont {Eichten}\ \emph {et~al.}(1975)\citenamefont {Eichten}, \citenamefont {Gottfried}, \citenamefont {Kinoshita}, \citenamefont {Kogut}, \citenamefont {Lane},\ and\ \citenamefont {Yan}}]{Eichten:1974af}%
  \BibitemOpen
  \bibfield  {author} {\bibinfo {author} {\bibfnamefont {E.}~\bibnamefont {Eichten}}, \bibinfo {author} {\bibfnamefont {K.}~\bibnamefont {Gottfried}}, \bibinfo {author} {\bibfnamefont {T.}~\bibnamefont {Kinoshita}}, \bibinfo {author} {\bibfnamefont {J.~B.}\ \bibnamefont {Kogut}}, \bibinfo {author} {\bibfnamefont {K.~D.}\ \bibnamefont {Lane}},\ and\ \bibinfo {author} {\bibfnamefont {T.-M.}\ \bibnamefont {Yan}},\ }\href {https://doi.org/10.1103/PhysRevLett.34.369} {\bibfield  {journal} {\bibinfo  {journal} {Phys. Rev. Lett.}\ }\textbf {\bibinfo {volume} {34}},\ \bibinfo {pages} {369} (\bibinfo {year} {1975})},\ \bibinfo {note} {[Erratum: Phys.Rev.Lett. 36, 1276 (1976)]}\BibitemShut {NoStop}%
\bibitem [{\citenamefont {Eichten}\ \emph {et~al.}(1978)\citenamefont {Eichten}, \citenamefont {Gottfried}, \citenamefont {Kinoshita}, \citenamefont {Lane},\ and\ \citenamefont {Yan}}]{Eichten:1978tg}%
  \BibitemOpen
  \bibfield  {author} {\bibinfo {author} {\bibfnamefont {E.}~\bibnamefont {Eichten}}, \bibinfo {author} {\bibfnamefont {K.}~\bibnamefont {Gottfried}}, \bibinfo {author} {\bibfnamefont {T.}~\bibnamefont {Kinoshita}}, \bibinfo {author} {\bibfnamefont {K.~D.}\ \bibnamefont {Lane}},\ and\ \bibinfo {author} {\bibfnamefont {T.-M.}\ \bibnamefont {Yan}},\ }\href {https://doi.org/10.1103/PhysRevD.17.3090} {\bibfield  {journal} {\bibinfo  {journal} {Phys. Rev. D}\ }\textbf {\bibinfo {volume} {17}},\ \bibinfo {pages} {3090} (\bibinfo {year} {1978})},\ \bibinfo {note} {[Erratum: Phys.Rev.D 21, 313 (1980)]}\BibitemShut {NoStop}%
\bibitem [{\citenamefont {Eichten}\ \emph {et~al.}(1980)\citenamefont {Eichten}, \citenamefont {Gottfried}, \citenamefont {Kinoshita}, \citenamefont {Lane},\ and\ \citenamefont {Yan}}]{Eichten:1979ms}%
  \BibitemOpen
  \bibfield  {author} {\bibinfo {author} {\bibfnamefont {E.}~\bibnamefont {Eichten}}, \bibinfo {author} {\bibfnamefont {K.}~\bibnamefont {Gottfried}}, \bibinfo {author} {\bibfnamefont {T.}~\bibnamefont {Kinoshita}}, \bibinfo {author} {\bibfnamefont {K.~D.}\ \bibnamefont {Lane}},\ and\ \bibinfo {author} {\bibfnamefont {T.-M.}\ \bibnamefont {Yan}},\ }\href {https://doi.org/10.1103/PhysRevD.21.203} {\bibfield  {journal} {\bibinfo  {journal} {Phys. Rev. D}\ }\textbf {\bibinfo {volume} {21}},\ \bibinfo {pages} {203} (\bibinfo {year} {1980})}\BibitemShut {NoStop}%
\bibitem [{\citenamefont {Buchmuller}\ and\ \citenamefont {Tye}(1981)}]{Buchmuller:1980su}%
  \BibitemOpen
  \bibfield  {author} {\bibinfo {author} {\bibfnamefont {W.}~\bibnamefont {Buchmuller}}\ and\ \bibinfo {author} {\bibfnamefont {S.~H.~H.}\ \bibnamefont {Tye}},\ }\href {https://doi.org/10.1103/PhysRevD.24.132} {\bibfield  {journal} {\bibinfo  {journal} {Phys. Rev. D}\ }\textbf {\bibinfo {volume} {24}},\ \bibinfo {pages} {132} (\bibinfo {year} {1981})}\BibitemShut {NoStop}%
\bibitem [{\citenamefont {Koma}\ \emph {et~al.}(2006)\citenamefont {Koma}, \citenamefont {Koma},\ and\ \citenamefont {Wittig}}]{Koma:2006si}%
  \BibitemOpen
  \bibfield  {author} {\bibinfo {author} {\bibfnamefont {Y.}~\bibnamefont {Koma}}, \bibinfo {author} {\bibfnamefont {M.}~\bibnamefont {Koma}},\ and\ \bibinfo {author} {\bibfnamefont {H.}~\bibnamefont {Wittig}},\ }\href {https://doi.org/10.1103/PhysRevLett.97.122003} {\bibfield  {journal} {\bibinfo  {journal} {Phys. Rev. Lett.}\ }\textbf {\bibinfo {volume} {97}},\ \bibinfo {pages} {122003} (\bibinfo {year} {2006})},\ \Eprint {https://arxiv.org/abs/hep-lat/0607009} {arXiv:hep-lat/0607009} \BibitemShut {NoStop}%
\bibitem [{\citenamefont {Brambilla}\ \emph {et~al.}(2005)\citenamefont {Brambilla}, \citenamefont {Pineda}, \citenamefont {Soto},\ and\ \citenamefont {Vairo}}]{Brambilla:2004jw}%
  \BibitemOpen
  \bibfield  {author} {\bibinfo {author} {\bibfnamefont {N.}~\bibnamefont {Brambilla}}, \bibinfo {author} {\bibfnamefont {A.}~\bibnamefont {Pineda}}, \bibinfo {author} {\bibfnamefont {J.}~\bibnamefont {Soto}},\ and\ \bibinfo {author} {\bibfnamefont {A.}~\bibnamefont {Vairo}},\ }\href {https://doi.org/10.1103/RevModPhys.77.1423} {\bibfield  {journal} {\bibinfo  {journal} {Rev. Mod. Phys.}\ }\textbf {\bibinfo {volume} {77}},\ \bibinfo {pages} {1423} (\bibinfo {year} {2005})},\ \Eprint {https://arxiv.org/abs/hep-ph/0410047} {arXiv:hep-ph/0410047} \BibitemShut {NoStop}%
\bibitem [{\citenamefont {Karsch}\ \emph {et~al.}(2006)\citenamefont {Karsch}, \citenamefont {Kharzeev},\ and\ \citenamefont {Satz}}]{Karsch:2005nk}%
  \BibitemOpen
  \bibfield  {author} {\bibinfo {author} {\bibfnamefont {F.}~\bibnamefont {Karsch}}, \bibinfo {author} {\bibfnamefont {D.}~\bibnamefont {Kharzeev}},\ and\ \bibinfo {author} {\bibfnamefont {H.}~\bibnamefont {Satz}},\ }\href {https://doi.org/10.1016/j.physletb.2006.03.078} {\bibfield  {journal} {\bibinfo  {journal} {Phys. Lett. B}\ }\textbf {\bibinfo {volume} {637}},\ \bibinfo {pages} {75} (\bibinfo {year} {2006})},\ \Eprint {https://arxiv.org/abs/hep-ph/0512239} {arXiv:hep-ph/0512239} \BibitemShut {NoStop}%
\bibitem [{\citenamefont {Chatrchyan}\ \emph {et~al.}(2011)\citenamefont {Chatrchyan} \emph {et~al.}}]{CMS:2011all}%
  \BibitemOpen
  \bibfield  {author} {\bibinfo {author} {\bibfnamefont {S.}~\bibnamefont {Chatrchyan}} \emph {et~al.} (\bibinfo {collaboration} {CMS}),\ }\href {https://doi.org/10.1103/PhysRevLett.107.052302} {\bibfield  {journal} {\bibinfo  {journal} {Phys. Rev. Lett.}\ }\textbf {\bibinfo {volume} {107}},\ \bibinfo {pages} {052302} (\bibinfo {year} {2011})},\ \Eprint {https://arxiv.org/abs/1105.4894} {arXiv:1105.4894 [nucl-ex]} \BibitemShut {NoStop}%
\bibitem [{\citenamefont {Chatrchyan}\ \emph {et~al.}(2012)\citenamefont {Chatrchyan} \emph {et~al.}}]{CMS:2012gvv}%
  \BibitemOpen
  \bibfield  {author} {\bibinfo {author} {\bibfnamefont {S.}~\bibnamefont {Chatrchyan}} \emph {et~al.} (\bibinfo {collaboration} {CMS}),\ }\href {https://doi.org/10.1103/PhysRevLett.109.222301} {\bibfield  {journal} {\bibinfo  {journal} {Phys. Rev. Lett.}\ }\textbf {\bibinfo {volume} {109}},\ \bibinfo {pages} {222301} (\bibinfo {year} {2012})},\ \bibinfo {note} {[Erratum: Phys.Rev.Lett. 120, 199903 (2018)]},\ \Eprint {https://arxiv.org/abs/1208.2826} {arXiv:1208.2826 [nucl-ex]} \BibitemShut {NoStop}%
\bibitem [{\citenamefont {Agotiya}\ \emph {et~al.}(2009)\citenamefont {Agotiya}, \citenamefont {Chandra},\ and\ \citenamefont {Patra}}]{Agotiya:2008ie}%
  \BibitemOpen
  \bibfield  {author} {\bibinfo {author} {\bibfnamefont {V.}~\bibnamefont {Agotiya}}, \bibinfo {author} {\bibfnamefont {V.}~\bibnamefont {Chandra}},\ and\ \bibinfo {author} {\bibfnamefont {B.~K.}\ \bibnamefont {Patra}},\ }\href {https://doi.org/10.1103/PhysRevC.80.025210} {\bibfield  {journal} {\bibinfo  {journal} {Phys. Rev. C}\ }\textbf {\bibinfo {volume} {80}},\ \bibinfo {pages} {025210} (\bibinfo {year} {2009})},\ \Eprint {https://arxiv.org/abs/0808.2699} {arXiv:0808.2699 [hep-ph]} \BibitemShut {NoStop}%
\bibitem [{\citenamefont {Romatschke}\ and\ \citenamefont {Strickland}(2003)}]{Romatschke:2003ms}%
  \BibitemOpen
  \bibfield  {author} {\bibinfo {author} {\bibfnamefont {P.}~\bibnamefont {Romatschke}}\ and\ \bibinfo {author} {\bibfnamefont {M.}~\bibnamefont {Strickland}},\ }\href {https://doi.org/10.1103/PhysRevD.68.036004} {\bibfield  {journal} {\bibinfo  {journal} {Phys. Rev. D}\ }\textbf {\bibinfo {volume} {68}},\ \bibinfo {pages} {036004} (\bibinfo {year} {2003})},\ \Eprint {https://arxiv.org/abs/hep-ph/0304092} {arXiv:hep-ph/0304092} \BibitemShut {NoStop}%
\bibitem [{\citenamefont {Romatschke}\ and\ \citenamefont {Strickland}(2004)}]{Romatschke:2004jh}%
  \BibitemOpen
  \bibfield  {author} {\bibinfo {author} {\bibfnamefont {P.}~\bibnamefont {Romatschke}}\ and\ \bibinfo {author} {\bibfnamefont {M.}~\bibnamefont {Strickland}},\ }\href {https://doi.org/10.1103/PhysRevD.70.116006} {\bibfield  {journal} {\bibinfo  {journal} {Phys. Rev. D}\ }\textbf {\bibinfo {volume} {70}},\ \bibinfo {pages} {116006} (\bibinfo {year} {2004})},\ \Eprint {https://arxiv.org/abs/hep-ph/0406188} {arXiv:hep-ph/0406188} \BibitemShut {NoStop}%
\bibitem [{\citenamefont {Mrowczynski}\ \emph {et~al.}(2004)\citenamefont {Mrowczynski}, \citenamefont {Rebhan},\ and\ \citenamefont {Strickland}}]{Mrowczynski:2004kv}%
  \BibitemOpen
  \bibfield  {author} {\bibinfo {author} {\bibfnamefont {S.}~\bibnamefont {Mrowczynski}}, \bibinfo {author} {\bibfnamefont {A.}~\bibnamefont {Rebhan}},\ and\ \bibinfo {author} {\bibfnamefont {M.}~\bibnamefont {Strickland}},\ }\href {https://doi.org/10.1103/PhysRevD.70.025004} {\bibfield  {journal} {\bibinfo  {journal} {Phys. Rev. D}\ }\textbf {\bibinfo {volume} {70}},\ \bibinfo {pages} {025004} (\bibinfo {year} {2004})},\ \Eprint {https://arxiv.org/abs/hep-ph/0403256} {arXiv:hep-ph/0403256} \BibitemShut {NoStop}%
\bibitem [{\citenamefont {Mrowczynski}\ \emph {et~al.}(2017)\citenamefont {Mrowczynski}, \citenamefont {Schenke},\ and\ \citenamefont {Strickland}}]{Mrowczynski:2016etf}%
  \BibitemOpen
  \bibfield  {author} {\bibinfo {author} {\bibfnamefont {S.}~\bibnamefont {Mrowczynski}}, \bibinfo {author} {\bibfnamefont {B.}~\bibnamefont {Schenke}},\ and\ \bibinfo {author} {\bibfnamefont {M.}~\bibnamefont {Strickland}},\ }\href {https://doi.org/10.1016/j.physrep.2017.03.003} {\bibfield  {journal} {\bibinfo  {journal} {Phys. Rept.}\ }\textbf {\bibinfo {volume} {682}},\ \bibinfo {pages} {1} (\bibinfo {year} {2017})},\ \Eprint {https://arxiv.org/abs/1603.08946} {arXiv:1603.08946 [hep-ph]} \BibitemShut {NoStop}%
\bibitem [{\citenamefont {Romatschke}\ and\ \citenamefont {Rebhan}(2006)}]{Romatschke:2006wg}%
  \BibitemOpen
  \bibfield  {author} {\bibinfo {author} {\bibfnamefont {P.}~\bibnamefont {Romatschke}}\ and\ \bibinfo {author} {\bibfnamefont {A.}~\bibnamefont {Rebhan}},\ }\href {https://doi.org/10.1103/PhysRevLett.97.252301} {\bibfield  {journal} {\bibinfo  {journal} {Phys. Rev. Lett.}\ }\textbf {\bibinfo {volume} {97}},\ \bibinfo {pages} {252301} (\bibinfo {year} {2006})},\ \Eprint {https://arxiv.org/abs/hep-ph/0605064} {arXiv:hep-ph/0605064} \BibitemShut {NoStop}%
\bibitem [{\citenamefont {Dumitru}\ \emph {et~al.}(2009{\natexlab{a}})\citenamefont {Dumitru}, \citenamefont {Guo},\ and\ \citenamefont {Strickland}}]{Dumitru:2009fy}%
  \BibitemOpen
  \bibfield  {author} {\bibinfo {author} {\bibfnamefont {A.}~\bibnamefont {Dumitru}}, \bibinfo {author} {\bibfnamefont {Y.}~\bibnamefont {Guo}},\ and\ \bibinfo {author} {\bibfnamefont {M.}~\bibnamefont {Strickland}},\ }\href {https://doi.org/10.1103/PhysRevD.79.114003} {\bibfield  {journal} {\bibinfo  {journal} {Phys. Rev. D}\ }\textbf {\bibinfo {volume} {79}},\ \bibinfo {pages} {114003} (\bibinfo {year} {2009}{\natexlab{a}})},\ \Eprint {https://arxiv.org/abs/0903.4703} {arXiv:0903.4703 [hep-ph]} \BibitemShut {NoStop}%
\bibitem [{\citenamefont {Karmakar}\ \emph {et~al.}(2022)\citenamefont {Karmakar}, \citenamefont {Ghosh},\ and\ \citenamefont {Mukherjee}}]{Karmakar:2022one}%
  \BibitemOpen
  \bibfield  {author} {\bibinfo {author} {\bibfnamefont {B.}~\bibnamefont {Karmakar}}, \bibinfo {author} {\bibfnamefont {R.}~\bibnamefont {Ghosh}},\ and\ \bibinfo {author} {\bibfnamefont {A.}~\bibnamefont {Mukherjee}},\ }\href {https://doi.org/10.1103/PhysRevD.106.116006} {\bibfield  {journal} {\bibinfo  {journal} {Phys. Rev. D}\ }\textbf {\bibinfo {volume} {106}},\ \bibinfo {pages} {116006} (\bibinfo {year} {2022})},\ \Eprint {https://arxiv.org/abs/2204.09646} {arXiv:2204.09646 [hep-ph]} \BibitemShut {NoStop}%
\bibitem [{\citenamefont {Dumitru}\ \emph {et~al.}(2008)\citenamefont {Dumitru}, \citenamefont {Guo},\ and\ \citenamefont {Strickland}}]{Dumitru:2007hy}%
  \BibitemOpen
  \bibfield  {author} {\bibinfo {author} {\bibfnamefont {A.}~\bibnamefont {Dumitru}}, \bibinfo {author} {\bibfnamefont {Y.}~\bibnamefont {Guo}},\ and\ \bibinfo {author} {\bibfnamefont {M.}~\bibnamefont {Strickland}},\ }\href {https://doi.org/10.1016/j.physletb.2008.02.048} {\bibfield  {journal} {\bibinfo  {journal} {Phys. Lett. B}\ }\textbf {\bibinfo {volume} {662}},\ \bibinfo {pages} {37} (\bibinfo {year} {2008})},\ \Eprint {https://arxiv.org/abs/0711.4722} {arXiv:0711.4722 [hep-ph]} \BibitemShut {NoStop}%
\bibitem [{\citenamefont {Burnier}\ \emph {et~al.}(2009)\citenamefont {Burnier}, \citenamefont {Laine},\ and\ \citenamefont {Vepsalainen}}]{Burnier:2009yu}%
  \BibitemOpen
  \bibfield  {author} {\bibinfo {author} {\bibfnamefont {Y.}~\bibnamefont {Burnier}}, \bibinfo {author} {\bibfnamefont {M.}~\bibnamefont {Laine}},\ and\ \bibinfo {author} {\bibfnamefont {M.}~\bibnamefont {Vepsalainen}},\ }\href {https://doi.org/10.1016/j.physletb.2009.05.067} {\bibfield  {journal} {\bibinfo  {journal} {Phys. Lett. B}\ }\textbf {\bibinfo {volume} {678}},\ \bibinfo {pages} {86} (\bibinfo {year} {2009})},\ \Eprint {https://arxiv.org/abs/0903.3467} {arXiv:0903.3467 [hep-ph]} \BibitemShut {NoStop}%
\bibitem [{\citenamefont {Dumitru}\ \emph {et~al.}(2009{\natexlab{b}})\citenamefont {Dumitru}, \citenamefont {Guo}, \citenamefont {Mocsy},\ and\ \citenamefont {Strickland}}]{Dumitru:2009ni}%
  \BibitemOpen
  \bibfield  {author} {\bibinfo {author} {\bibfnamefont {A.}~\bibnamefont {Dumitru}}, \bibinfo {author} {\bibfnamefont {Y.}~\bibnamefont {Guo}}, \bibinfo {author} {\bibfnamefont {A.}~\bibnamefont {Mocsy}},\ and\ \bibinfo {author} {\bibfnamefont {M.}~\bibnamefont {Strickland}},\ }\href {https://doi.org/10.1103/PhysRevD.79.054019} {\bibfield  {journal} {\bibinfo  {journal} {Phys. Rev. D}\ }\textbf {\bibinfo {volume} {79}},\ \bibinfo {pages} {054019} (\bibinfo {year} {2009}{\natexlab{b}})},\ \Eprint {https://arxiv.org/abs/0901.1998} {arXiv:0901.1998 [hep-ph]} \BibitemShut {NoStop}%
\bibitem [{\citenamefont {Thakur}\ \emph {et~al.}(2013)\citenamefont {Thakur}, \citenamefont {Haque}, \citenamefont {Kakade},\ and\ \citenamefont {Patra}}]{Thakur:2012eb}%
  \BibitemOpen
  \bibfield  {author} {\bibinfo {author} {\bibfnamefont {L.}~\bibnamefont {Thakur}}, \bibinfo {author} {\bibfnamefont {N.}~\bibnamefont {Haque}}, \bibinfo {author} {\bibfnamefont {U.}~\bibnamefont {Kakade}},\ and\ \bibinfo {author} {\bibfnamefont {B.~K.}\ \bibnamefont {Patra}},\ }\href {https://doi.org/10.1103/PhysRevD.88.054022} {\bibfield  {journal} {\bibinfo  {journal} {Phys. Rev. D}\ }\textbf {\bibinfo {volume} {88}},\ \bibinfo {pages} {054022} (\bibinfo {year} {2013})},\ \Eprint {https://arxiv.org/abs/1212.2803} {arXiv:1212.2803 [hep-ph]} \BibitemShut {NoStop}%
\bibitem [{\citenamefont {Dong}\ \emph {et~al.}(2022)\citenamefont {Dong}, \citenamefont {Guo}, \citenamefont {Islam}, \citenamefont {Rothkopf},\ and\ \citenamefont {Strickland}}]{Dong:2022mbo}%
  \BibitemOpen
  \bibfield  {author} {\bibinfo {author} {\bibfnamefont {L.}~\bibnamefont {Dong}}, \bibinfo {author} {\bibfnamefont {Y.}~\bibnamefont {Guo}}, \bibinfo {author} {\bibfnamefont {A.}~\bibnamefont {Islam}}, \bibinfo {author} {\bibfnamefont {A.}~\bibnamefont {Rothkopf}},\ and\ \bibinfo {author} {\bibfnamefont {M.}~\bibnamefont {Strickland}},\ }\href {https://doi.org/10.1007/JHEP09(2022)200} {\bibfield  {journal} {\bibinfo  {journal} {JHEP}\ }\textbf {\bibinfo {volume} {09}},\ \bibinfo {pages} {200}},\ \Eprint {https://arxiv.org/abs/2205.10349} {arXiv:2205.10349 [hep-ph]} \BibitemShut {NoStop}%
\bibitem [{\citenamefont {Ghosh}\ \emph {et~al.}(2022)\citenamefont {Ghosh}, \citenamefont {Bandyopadhyay}, \citenamefont {Nilima},\ and\ \citenamefont {Ghosh}}]{Ghosh:2022sxi}%
  \BibitemOpen
  \bibfield  {author} {\bibinfo {author} {\bibfnamefont {R.}~\bibnamefont {Ghosh}}, \bibinfo {author} {\bibfnamefont {A.}~\bibnamefont {Bandyopadhyay}}, \bibinfo {author} {\bibfnamefont {I.}~\bibnamefont {Nilima}},\ and\ \bibinfo {author} {\bibfnamefont {S.}~\bibnamefont {Ghosh}},\ }\href {https://doi.org/10.1103/PhysRevD.106.054010} {\bibfield  {journal} {\bibinfo  {journal} {Phys. Rev. D}\ }\textbf {\bibinfo {volume} {106}},\ \bibinfo {pages} {054010} (\bibinfo {year} {2022})},\ \Eprint {https://arxiv.org/abs/2204.02312} {arXiv:2204.02312 [hep-ph]} \BibitemShut {NoStop}%
\bibitem [{\citenamefont {Carrington}\ \emph {et~al.}(2024)\citenamefont {Carrington}, \citenamefont {Kunstatter},\ and\ \citenamefont {Mukherjee}}]{Carrington:2024rxw}%
  \BibitemOpen
  \bibfield  {author} {\bibinfo {author} {\bibfnamefont {M.~E.}\ \bibnamefont {Carrington}}, \bibinfo {author} {\bibfnamefont {G.}~\bibnamefont {Kunstatter}},\ and\ \bibinfo {author} {\bibfnamefont {A.}~\bibnamefont {Mukherjee}},\ }\href {https://doi.org/10.1103/PhysRevC.110.035203} {\bibfield  {journal} {\bibinfo  {journal} {Phys. Rev. C}\ }\textbf {\bibinfo {volume} {110}},\ \bibinfo {pages} {035203} (\bibinfo {year} {2024})},\ \Eprint {https://arxiv.org/abs/2405.05622} {arXiv:2405.05622 [hep-ph]} \BibitemShut {NoStop}%
\bibitem [{\citenamefont {Zhao}\ \emph {et~al.}(2023)\citenamefont {Zhao}, \citenamefont {Qiu}, \citenamefont {Guo},\ and\ \citenamefont {Strickland}}]{Zhao:2023mrz}%
  \BibitemOpen
  \bibfield  {author} {\bibinfo {author} {\bibfnamefont {R.}~\bibnamefont {Zhao}}, \bibinfo {author} {\bibfnamefont {L.}~\bibnamefont {Qiu}}, \bibinfo {author} {\bibfnamefont {Y.}~\bibnamefont {Guo}},\ and\ \bibinfo {author} {\bibfnamefont {M.}~\bibnamefont {Strickland}},\ }\href {https://doi.org/10.1103/PhysRevD.108.034023} {\bibfield  {journal} {\bibinfo  {journal} {Phys. Rev. D}\ }\textbf {\bibinfo {volume} {108}},\ \bibinfo {pages} {034023} (\bibinfo {year} {2023})},\ \Eprint {https://arxiv.org/abs/2306.12851} {arXiv:2306.12851 [hep-ph]} \BibitemShut {NoStop}%
\bibitem [{\citenamefont {Carrington}\ \emph {et~al.}(2004)\citenamefont {Carrington}, \citenamefont {Fugleberg}, \citenamefont {Pickering},\ and\ \citenamefont {Thoma}}]{Carrington:2003je}%
  \BibitemOpen
  \bibfield  {author} {\bibinfo {author} {\bibfnamefont {M.~E.}\ \bibnamefont {Carrington}}, \bibinfo {author} {\bibfnamefont {T.}~\bibnamefont {Fugleberg}}, \bibinfo {author} {\bibfnamefont {D.}~\bibnamefont {Pickering}},\ and\ \bibinfo {author} {\bibfnamefont {M.~H.}\ \bibnamefont {Thoma}},\ }\href {https://doi.org/10.1139/p04-035} {\bibfield  {journal} {\bibinfo  {journal} {Can. J. Phys.}\ }\textbf {\bibinfo {volume} {82}},\ \bibinfo {pages} {671} (\bibinfo {year} {2004})},\ \Eprint {https://arxiv.org/abs/hep-ph/0312103} {arXiv:hep-ph/0312103} \BibitemShut {NoStop}%
\bibitem [{\citenamefont {Schenke}\ \emph {et~al.}(2006)\citenamefont {Schenke}, \citenamefont {Strickland}, \citenamefont {Greiner},\ and\ \citenamefont {Thoma}}]{Schenke:2006xu}%
  \BibitemOpen
  \bibfield  {author} {\bibinfo {author} {\bibfnamefont {B.}~\bibnamefont {Schenke}}, \bibinfo {author} {\bibfnamefont {M.}~\bibnamefont {Strickland}}, \bibinfo {author} {\bibfnamefont {C.}~\bibnamefont {Greiner}},\ and\ \bibinfo {author} {\bibfnamefont {M.~H.}\ \bibnamefont {Thoma}},\ }\href {https://doi.org/10.1103/PhysRevD.73.125004} {\bibfield  {journal} {\bibinfo  {journal} {Phys. Rev. D}\ }\textbf {\bibinfo {volume} {73}},\ \bibinfo {pages} {125004} (\bibinfo {year} {2006})},\ \Eprint {https://arxiv.org/abs/hep-ph/0603029} {arXiv:hep-ph/0603029} \BibitemShut {NoStop}%
\bibitem [{\citenamefont {Kumar}\ \emph {et~al.}(2018)\citenamefont {Kumar}, \citenamefont {Jamal}, \citenamefont {Chandra},\ and\ \citenamefont {Bhatt}}]{Kumar:2017bja}%
  \BibitemOpen
  \bibfield  {author} {\bibinfo {author} {\bibfnamefont {A.}~\bibnamefont {Kumar}}, \bibinfo {author} {\bibfnamefont {M.~Y.}\ \bibnamefont {Jamal}}, \bibinfo {author} {\bibfnamefont {V.}~\bibnamefont {Chandra}},\ and\ \bibinfo {author} {\bibfnamefont {J.~R.}\ \bibnamefont {Bhatt}},\ }\href {https://doi.org/10.1103/PhysRevD.97.034007} {\bibfield  {journal} {\bibinfo  {journal} {Phys. Rev. D}\ }\textbf {\bibinfo {volume} {97}},\ \bibinfo {pages} {034007} (\bibinfo {year} {2018})},\ \Eprint {https://arxiv.org/abs/1709.01032} {arXiv:1709.01032 [nucl-th]} \BibitemShut {NoStop}%
\bibitem [{\citenamefont {Thakur}\ \emph {et~al.}(2014)\citenamefont {Thakur}, \citenamefont {Kakade},\ and\ \citenamefont {Patra}}]{Thakur:2013nia}%
  \BibitemOpen
  \bibfield  {author} {\bibinfo {author} {\bibfnamefont {L.}~\bibnamefont {Thakur}}, \bibinfo {author} {\bibfnamefont {U.}~\bibnamefont {Kakade}},\ and\ \bibinfo {author} {\bibfnamefont {B.~K.}\ \bibnamefont {Patra}},\ }\href {https://doi.org/10.1103/PhysRevD.89.094020} {\bibfield  {journal} {\bibinfo  {journal} {Phys. Rev. D}\ }\textbf {\bibinfo {volume} {89}},\ \bibinfo {pages} {094020} (\bibinfo {year} {2014})},\ \Eprint {https://arxiv.org/abs/1401.0172} {arXiv:1401.0172 [hep-ph]} \BibitemShut {NoStop}%
\bibitem [{\citenamefont {Nopoush}\ \emph {et~al.}(2017)\citenamefont {Nopoush}, \citenamefont {Guo},\ and\ \citenamefont {Strickland}}]{Nopoush:2017zbu}%
  \BibitemOpen
  \bibfield  {author} {\bibinfo {author} {\bibfnamefont {M.}~\bibnamefont {Nopoush}}, \bibinfo {author} {\bibfnamefont {Y.}~\bibnamefont {Guo}},\ and\ \bibinfo {author} {\bibfnamefont {M.}~\bibnamefont {Strickland}},\ }\href {https://doi.org/10.1007/JHEP09(2017)063} {\bibfield  {journal} {\bibinfo  {journal} {JHEP}\ }\textbf {\bibinfo {volume} {09}},\ \bibinfo {pages} {063}},\ \Eprint {https://arxiv.org/abs/1706.08091} {arXiv:1706.08091 [hep-ph]} \BibitemShut {NoStop}%
\bibitem [{\citenamefont {Carrington}\ \emph {et~al.}(1999)\citenamefont {Carrington}, \citenamefont {Hou},\ and\ \citenamefont {Thoma}}]{Carrington:1997sq}%
  \BibitemOpen
  \bibfield  {author} {\bibinfo {author} {\bibfnamefont {M.~E.}\ \bibnamefont {Carrington}}, \bibinfo {author} {\bibfnamefont {D.-f.}\ \bibnamefont {Hou}},\ and\ \bibinfo {author} {\bibfnamefont {M.~H.}\ \bibnamefont {Thoma}},\ }\href {https://doi.org/10.1007/s100520050412} {\bibfield  {journal} {\bibinfo  {journal} {Eur. Phys. J. C}\ }\textbf {\bibinfo {volume} {7}},\ \bibinfo {pages} {347} (\bibinfo {year} {1999})},\ \Eprint {https://arxiv.org/abs/hep-ph/9708363} {arXiv:hep-ph/9708363} \BibitemShut {NoStop}%
\bibitem [{\citenamefont {Sebastian}\ \emph {et~al.}(2023{\natexlab{a}})\citenamefont {Sebastian}, \citenamefont {Thakur}, \citenamefont {Mishra},\ and\ \citenamefont {Haque}}]{Sebastian:2023tlw}%
  \BibitemOpen
  \bibfield  {author} {\bibinfo {author} {\bibfnamefont {J.}~\bibnamefont {Sebastian}}, \bibinfo {author} {\bibfnamefont {L.}~\bibnamefont {Thakur}}, \bibinfo {author} {\bibfnamefont {H.}~\bibnamefont {Mishra}},\ and\ \bibinfo {author} {\bibfnamefont {N.}~\bibnamefont {Haque}},\ }\href {https://doi.org/10.1103/PhysRevD.108.094001} {\bibfield  {journal} {\bibinfo  {journal} {Phys. Rev. D}\ }\textbf {\bibinfo {volume} {108}},\ \bibinfo {pages} {094001} (\bibinfo {year} {2023}{\natexlab{a}})},\ \Eprint {https://arxiv.org/abs/2308.04410} {arXiv:2308.04410 [hep-ph]} \BibitemShut {NoStop}%
\bibitem [{\citenamefont {Singh}\ \emph {et~al.}(2018)\citenamefont {Singh}, \citenamefont {Thakur},\ and\ \citenamefont {Mishra}}]{Singh:2017nfa}%
  \BibitemOpen
  \bibfield  {author} {\bibinfo {author} {\bibfnamefont {B.}~\bibnamefont {Singh}}, \bibinfo {author} {\bibfnamefont {L.}~\bibnamefont {Thakur}},\ and\ \bibinfo {author} {\bibfnamefont {H.}~\bibnamefont {Mishra}},\ }\href {https://doi.org/10.1103/PhysRevD.97.096011} {\bibfield  {journal} {\bibinfo  {journal} {Phys. Rev. D}\ }\textbf {\bibinfo {volume} {97}},\ \bibinfo {pages} {096011} (\bibinfo {year} {2018})},\ \Eprint {https://arxiv.org/abs/1711.03071} {arXiv:1711.03071 [hep-ph]} \BibitemShut {NoStop}%
\bibitem [{\citenamefont {Thakur}\ \emph {et~al.}(2017)\citenamefont {Thakur}, \citenamefont {Haque},\ and\ \citenamefont {Mishra}}]{Thakur:2016cki}%
  \BibitemOpen
  \bibfield  {author} {\bibinfo {author} {\bibfnamefont {L.}~\bibnamefont {Thakur}}, \bibinfo {author} {\bibfnamefont {N.}~\bibnamefont {Haque}},\ and\ \bibinfo {author} {\bibfnamefont {H.}~\bibnamefont {Mishra}},\ }\href {https://doi.org/10.1103/PhysRevD.95.036014} {\bibfield  {journal} {\bibinfo  {journal} {Phys. Rev. D}\ }\textbf {\bibinfo {volume} {95}},\ \bibinfo {pages} {036014} (\bibinfo {year} {2017})},\ \Eprint {https://arxiv.org/abs/1611.04568} {arXiv:1611.04568 [hep-ph]} \BibitemShut {NoStop}%
\bibitem [{\citenamefont {Sebastian}\ \emph {et~al.}(2023{\natexlab{b}})\citenamefont {Sebastian}, \citenamefont {Jamal},\ and\ \citenamefont {Haque}}]{Sebastian:2022sga}%
  \BibitemOpen
  \bibfield  {author} {\bibinfo {author} {\bibfnamefont {J.}~\bibnamefont {Sebastian}}, \bibinfo {author} {\bibfnamefont {M.~Y.}\ \bibnamefont {Jamal}},\ and\ \bibinfo {author} {\bibfnamefont {N.}~\bibnamefont {Haque}},\ }\href {https://doi.org/10.1103/PhysRevD.107.054040} {\bibfield  {journal} {\bibinfo  {journal} {Phys. Rev. D}\ }\textbf {\bibinfo {volume} {107}},\ \bibinfo {pages} {054040} (\bibinfo {year} {2023}{\natexlab{b}})},\ \Eprint {https://arxiv.org/abs/2207.08510} {arXiv:2207.08510 [hep-ph]} \BibitemShut {NoStop}%
\bibitem [{\citenamefont {Du}\ \emph {et~al.}(2017)\citenamefont {Du}, \citenamefont {Dumitru}, \citenamefont {Guo},\ and\ \citenamefont {Strickland}}]{Du:2016wdx}%
  \BibitemOpen
  \bibfield  {author} {\bibinfo {author} {\bibfnamefont {Q.}~\bibnamefont {Du}}, \bibinfo {author} {\bibfnamefont {A.}~\bibnamefont {Dumitru}}, \bibinfo {author} {\bibfnamefont {Y.}~\bibnamefont {Guo}},\ and\ \bibinfo {author} {\bibfnamefont {M.}~\bibnamefont {Strickland}},\ }\href {https://doi.org/10.1007/JHEP01(2017)123} {\bibfield  {journal} {\bibinfo  {journal} {JHEP}\ }\textbf {\bibinfo {volume} {01}},\ \bibinfo {pages} {123}},\ \Eprint {https://arxiv.org/abs/1611.08379} {arXiv:1611.08379 [hep-ph]} \BibitemShut {NoStop}%
\bibitem [{\citenamefont {Thakur}\ \emph {et~al.}(2020)\citenamefont {Thakur}, \citenamefont {Haque},\ and\ \citenamefont {Hirono}}]{Thakur:2020ifi}%
  \BibitemOpen
  \bibfield  {author} {\bibinfo {author} {\bibfnamefont {L.}~\bibnamefont {Thakur}}, \bibinfo {author} {\bibfnamefont {N.}~\bibnamefont {Haque}},\ and\ \bibinfo {author} {\bibfnamefont {Y.}~\bibnamefont {Hirono}},\ }\href {https://doi.org/10.1007/JHEP06(2020)071} {\bibfield  {journal} {\bibinfo  {journal} {JHEP}\ }\textbf {\bibinfo {volume} {06}},\ \bibinfo {pages} {071}},\ \Eprint {https://arxiv.org/abs/2004.03426} {arXiv:2004.03426 [hep-ph]} \BibitemShut {NoStop}%
\bibitem [{\citenamefont {Thakur}\ and\ \citenamefont {Hirono}(2022)}]{Thakur:2021vbo}%
  \BibitemOpen
  \bibfield  {author} {\bibinfo {author} {\bibfnamefont {L.}~\bibnamefont {Thakur}}\ and\ \bibinfo {author} {\bibfnamefont {Y.}~\bibnamefont {Hirono}},\ }\href {https://doi.org/10.1007/JHEP02(2022)207} {\bibfield  {journal} {\bibinfo  {journal} {JHEP}\ }\textbf {\bibinfo {volume} {02}},\ \bibinfo {pages} {207}},\ \Eprint {https://arxiv.org/abs/2111.08225} {arXiv:2111.08225 [hep-ph]} \BibitemShut {NoStop}%
\bibitem [{\citenamefont {Debnath}\ \emph {et~al.}(2024)\citenamefont {Debnath}, \citenamefont {Ghosh},\ and\ \citenamefont {Haque}}]{Debnath:2023dhs}%
  \BibitemOpen
  \bibfield  {author} {\bibinfo {author} {\bibfnamefont {M.}~\bibnamefont {Debnath}}, \bibinfo {author} {\bibfnamefont {R.}~\bibnamefont {Ghosh}},\ and\ \bibinfo {author} {\bibfnamefont {N.}~\bibnamefont {Haque}},\ }\href {https://doi.org/10.1140/epjc/s10052-024-12656-2} {\bibfield  {journal} {\bibinfo  {journal} {Eur. Phys. J. C}\ }\textbf {\bibinfo {volume} {84}},\ \bibinfo {pages} {313} (\bibinfo {year} {2024})},\ \Eprint {https://arxiv.org/abs/2305.16250} {arXiv:2305.16250 [hep-ph]} \BibitemShut {NoStop}%
\bibitem [{\citenamefont {Wang}\ and\ \citenamefont {Shovkovy}(2021)}]{Wang:2021ebh}%
  \BibitemOpen
  \bibfield  {author} {\bibinfo {author} {\bibfnamefont {X.}~\bibnamefont {Wang}}\ and\ \bibinfo {author} {\bibfnamefont {I.}~\bibnamefont {Shovkovy}},\ }\href {https://doi.org/10.1103/PhysRevD.104.056017} {\bibfield  {journal} {\bibinfo  {journal} {Phys. Rev. D}\ }\textbf {\bibinfo {volume} {104}},\ \bibinfo {pages} {056017} (\bibinfo {year} {2021})},\ \Eprint {https://arxiv.org/abs/2103.01967} {arXiv:2103.01967 [nucl-th]} \BibitemShut {NoStop}%
\end{thebibliography}

%

\end{document}